\documentclass[aps,prd,longbibliography, amsmath,amssymb,amsfonts,superscriptaddress,floatfix,nofootinbib]{revtex4-1} 
\pdfoutput=1
\usepackage{amsmath,amsthm,amsfonts,amssymb}
\usepackage{graphicx}

\usepackage{color}
\usepackage{hyperref}
\usepackage{algorithm}
\usepackage[capitalize,nameinlink]{cleveref}

\newcommand{\be}{\begin{equation}}
\newcommand{\ee}{\end{equation}}

\newcommand{\Z}{\mathbb{Z}}

\newcommand{\veck}{{\vec{k}}}
\newcommand{\vecr}{{\vec{r}}}
\newcommand{\hmu}{{\hat{\mu}}}
\newcommand{\hnu}{{\hat{\nu}}}
\newcommand{\bgamma}{{\bar{\gamma}}}

\newcommand{\talpha}{{\tilde{\alpha}}}
\newcommand{\tU}{{\tilde{U}}}
\newcommand{\fl}{{\text{fl}}}
\newcommand{\lay}{{\text{lay}}}
\newcommand{\vk}{{\vec{k}}_{2\text{d}}}

\newcommand{\Un}{U^{\text{nucl}}}
\newcommand{\Ua}{U^{\text{annih}}}

\newcommand{\Ufr}{U_{\text{framing}}}

\begin{document}
\title{Pumping Chirality in Three Dimensions}
\author{Lukasz Fidkowski}
\affiliation{Department of Physics, University of Washington, Seattle WA USA 98195-1560}
\author{Matthew B.~Hastings}
\affiliation{Station Q, Microsoft Research, Santa Barbara, CA USA 93106-6105}
\affiliation{Microsoft Quantum and Microsoft Research, Redmond, WA USA 98052}
\begin{abstract}
Using bosonization, which maps fermions coupled to a $\Z_2$ gauge field to a qubit system, we give a simple form for the three fermion non-trivial quantum cellular automaton (QCA) as realizing a phase depending on the \emph{framing} of flux loops.
We relate this framing dependent phase to a pump of $8$ copies of a $p+ip$ state through the system.  We give a resolution of an apparent paradox, namely that the pump is a shallow depth circuit (albeit with tails), while the QCA is nontrivial.  We discuss also the pump of fewer copies of a $p+ip$ state, and describe its action on topologically degenerate ground states.  One consequence of our results is that a pump of $n$ $p+ip$ states generated by a free fermi evolution is a free fermion unitary characterized by a non-trivial winding number $n$ as a map from the third homotopy group of the Brilliouin Zone $3$-torus to that of $SU(N_ b)$, where $N_b$ is the number of bands.  Using our simplified form of the QCA, we give higher dimensional generalizations that we conjecture are also nontrivial QCAs, and we discuss the relation to Chern-Simons theory.
\end{abstract}
\maketitle

\section{Introduction}

In recent years tools from quantum information theory have shown to be useful in understanding and classifying topological and symmetry protected quantum phases of matter.  One particularly important notion is that of local unitary equivalence, with two states of a many-qubit Hilbert space being equivalent if one can be obtained from the other, at least approximately, by applying a shallow depth circuit of local unitary operators.  Here `shallow' means that the depth of the circuit does not grow with system size in the thermodynamic limit.  This is an equivalence relation on short-range entangled states and allows one to define phases; when the gates of the circuit respect an underlying symmetry, the resulting phases are symmetry-protected.  In particular, a symmetry protected topological (SPT) phase is non-trivial if it can be disentangled into a symmetric tensor product state using a shallow depth circuit, but only at the expense of at least some of the gates in the circuit being non-symmetric.

A large class of SPT phases can be obtained by using the group cohomology construction \cite{Chen2013}.  For these in-cohomology phases, exact zero correlation length Hamiltonian models can be written down explicitly, as can their corresponding disentangling circuits.  These circuits have the special property that they are symmetric as unitary operators (despite some of their gates not being symmetric).  Soon after the discovery of the in-cohomology SPT phases it became clear that the cohomology construction does not capture all possible SPT phases.  One of the earliest examples of such a beyond-cohomology SPT phase is that given by the three-fermion Walker-Wang (3F WW) model, which defines a $3$ dimensional bosonic SPT phase of time reversal symmetry \cite{BCFV}.  This model is interesting because despite its seemingly simple structure - it is a Pauli stabilizer Hamiltonian - the construction of an explicit ground state disentangler has proved to be elusive, despite heuristic arguments showing that such a shallow circuit disentangler should certainly exist.

Despite this difficulty, it turns out that one can define an automorphism of the local operator algebra that disentangles the 3F WW Hamiltonian into that of a product state \cite{QCA}.  Such an automorphism must be implemented by some unitary that maps local operators to nearby local operators.  However, it was shown in \cite{QCA} that such a unitary cannot be a shallow depth circuit; assuming otherwise would lead to a contradiction, namely the existence of a standalone commuting projector model for the three-fermion topological order which lives at the boundary of this phase \cite{Kitaev_2005}.  Therefore, this unitary is an example of a quantum cellular automaton (QCA): it is a locality-preserving unitary that cannot be locally generated as a shallow depth circuit.

Any two-dimensional realization of the three-fermion topological order is necessarily chiral, i.e. has chiral transport of energy along its edge \cite{Kitaev_2005}.  One can construct an alternative boundary for the 3F WW model by condensing its three-fermion topological order against that of a standalone two-dimensional realization made out of ancilla quibits living on the boundary (namely, one condenses bosons which are bound states of the fermions in the two copies).  This alternative boundary has no topological order, but it is chiral.  Now, the three-fermion QCA, or rather its inverse, has the property that it creates a three-fermion Walker-Wang (3F WW) model from a trivial product state.  Hence, in a certain sense, it pumps chirality out to the boundary.

In this paper we make this idea of `pumping chirality' more precise.  First, we find a simple explicit form for the 3F QCA as a unitary operator\footnote{More precisely, the QCA is conjugation by this unitary, but for brevity we will often say that a QCA {\emph is} some unitary if it is realized by conjugation by that unitary.}.  Namely, by building off the simplified form of this 3F QCA introduced in \cite{Shirley}, we find another, circuit-equivalent\footnote{That is, equivalent up to conjugation by some low-depth quantum circuit.} form of the 3F QCA whose unitary action $U$ is diagonalized by writing the $3d$ bosonic spin Hilbert space as the Hilbert space of a fermion coupled to a $\Z_2$ gauge field.  Then any state $|\{f\},\{L\}\rangle$, where $\{f\}$ is a configuration of fermions and $\{L\}$ a configuration of $\Z_2$ gauge flux loops, is an eigenstate of $U$ with eigenvalue $(-1)^{\text{framing}(L)}$, where $\text{framing(L)}$ is the \emph{framing}\footnote{A framing is an assignment of $0$ or $1$ to framed curves (i.e., curves with a choice of frame for normal vectors), which changes by $1 \mod 2$ when the frame is twisted by $2\pi$.  Note that we assign $0$ or $1$ rather than $\pm 1$, as some authors do.  We use a particular assignment of framing to curves, called the blackboard framing, which then
assigns $0$ or $1$ to any given curve.  The precise definition of framing for us is given in \cref{review1f}.}
 of $L$.  This uniquely defines the action of the QCA.  This exact description of the QCA at the level of a unitary operator, rather than just an opaque description as an algebra automorphism, is one of the contributions of this paper.

Consider now a different unitary operator $U'$, which creates a bubble of $8$ copies of a $p+ip$ state (coupled to the $3d$ $\Z_2$ gauge field), pumps it through the entire $3d$ system, and annihilates it.  We explicitly construct such an operator below; it can be thought of as a 3 dimensional generalization of the Thouless pump, except instead of a 0d charge pumped across a 1d system, now a $2d$ chiral state is pumped across a $3d$ system.  This operator $U'$ should be thought of as a generalized symmetry, in the sense of the recent classification of such generalized symmetry actions by gauged invertible defects \cite{Barkeshli_2023}.  $U'$ is interesting for our purposes because it seems to have a similar action on the Hilbert space as the QCA $U$.  Namely, in the context of the pumping process defining $U'$, a configuration of $\Z_2$ gauge flux loops $\{L\}$ can be viewed as a braiding history of a magnetic flux particle; the sweeping process then simply computes the braiding phase associated with this configuration, and it is well known that in $8$ copies of the $p+ip$ state the magnetic flux particle is a fermion, so this braiding phase should be $(-1)^{\text{framing}(L)}$ \cite{Kitaev_2005}.  This is exactly the same as the action of the QCA $U$, as we just discussed.

However, this results in an apparent paradox, since we know that $U'$, by virtue of its definition, is an approximate shallow depth circuit, whereas the 3F QCA $U$ cannot be a such a circuit.  We find that the resolution is the following: the two operators only agree on the ``no-fermion" subspace of the Hilbert space.  This is the subspace where all the fermion occupation numbers are $0$, but the gauge flux configurations are arbitrary.  However, away from this ``no-fermion'' subspace, $U$ and $U'$ are different.  Indeed, whereas the 3F QCA $U$ acts as the identity on any configuration of fermions in the no flux sector, we show that the $8(p+ip)$ pumping operator must act non-trivially away from the "no-fermion" subspace.

The fact that pumping any number of $p+ip$ states must have a non-trivial action away from the ``no-fermion'' subspace even while preserving the ``no-fermion'' subspace has a simple free fermion interpretation that may be interesting in its own right.  Namely, in a {\emph{fermionic}} lattice Hilbert space, the unitary operator $U^{\text{p+ip}}$ that pumps a single copy of a $p+ip$ state can be constructed in a translationally invariant two band model, and characterized by the fact that over the Brillouin zone the unitaries $U_{\veck}^{\text{p+ip}}$, viewed as a map from $T^3$ to $SU(2)$ has a non-trivial winding number, or degree (this relies on the fact that $\pi_3(SU(2))=\Z$).  In particular, it cannot be the identity operator, in contrast to the usual one dimensional Thouless pump.  The non-zero quantized winding number also means that $U^{\text{p+ip}}$ cannot be a shallow circuit of particle-number conserving fermions; however, we show that it is a shallow circuit of $\Z_2$ conserved fermions and indeed such a circuit just implements a pump of a $p+ip$ state.  This construction answers a question posed in \cite{Rudner_pump}, where this winding number was identified (see eq. $4$ in \cite{Rudner_pump}) but no example saturating it was given.

%[Conjectural:] We also make a connection between our simplified QCA and the notion of pumping chirality in quantum field theory.  Namely, we consider a $3+1$d model of $4$ Dirac fermions coupled to a non-abelian gauge field ($SU(4)$?).  Since the fermions carry gauge charge, this can be emerged from a bosonic lattice spin model.  Including a mass term for all the Dirac fermions results in a trivial gapped confining gauge theory at low energies, corresponding to short range entangled (SRE) state.  Both all positive and all negative mass terms correspond to such SRE states, and furthermore are time reversal symmetric [should figure out how time reversal acts on the fermions]; however, they differ by the beyond cohomology time reversal SPT state of bosons, by the usual chirality arguments.  Thus, a $\pi$ chiral rotation of all $4$ fermions should on the lattice correspond to our non-trivial QCA action.  We show this by connecting such a $\pi$ chiral rotation to the statistical Witten effect in the field theory.  Once we Higgs such a theory down to $(-1)^F\subset SU(4)$ we get a fermionic $\Z_2$ gauge theory, and the flux loop excitations in this theory can be thought of as monopole worldlines (even though the theory has no monopoles - need to find a way to explain this).  The $(-1)^{\text{framing}}$ sign associated to each such flux loop is then a reflection of the fact that a $\pi$ chiral rotation turns such monopoles from bosonic to fermionic [need to check this].

If we pump fewer than 8 copies of a $p+ip$ state, then the gauge flux loop configuration $L$ describes the braiding history of a particle with some other statistics.  In the case of pumping a single copy of $p+ip$, a magnetic flux particle behaves as a $\sigma$ particle in the Ising TQFT, and so is a non-abelian anyon.  This is consistent with the fact that a magnetic vortex in a two-dimensional $p+ip$ state binds a zero mode.  In this case, the effect of braiding is more complicated, and if flux loops link with each other then the pump may transfer charge from one to the other.  However, even in the zero flux state, the pump can still have a nontrivial effect if the ground state of the system has a topological degeneracy as we discuss in \cref{backact}.  Indeed, this gives a way to realize a non-Clifford CCZ gate in a three-dimensional theory \cite{Barkeshli_2023}.

The rest of this paper is structured as follows.  In \cref{backact}, we begin with the case of vanishing gauge flux and the fermions in the ground state of a trivial atomic insulator.  We construct a free fermion quasi-local Hermitian operator whose evolution pumps a $p+ip$ paired state across the system.  Even though the endpoint of this evolution is described by some unitary operator that preserves the atomic insulator ground state, we show that this unitary operator cannot be the identity, and is in fact characterized by a non-trivial winding number.  We then show that, in the presence of a dynamical $\Z_2^f$ (fermion parity) gauge field, this operator implements a CCZ gate on the topologically degenerate ground state subspace of a spatial $3$-torus.  In \cref{multicopypump}, we consider the case where the fermions are in the ground state, but there may be non-zero gauge flux, and consider pumping different numbers of copies of $p+ip$ states.
In \cref{simpleqca}, we review the construction of the 3F QCA from \cite{Shirley} and find a simple expression for the action of this QCA as a unitary operator.  We also construct an even simpler unitary which is circuit equivalent to it, and show explicitly that these unitaries trivialize in the presence of fundamental fermions.  In \cref{semionQCA}, we construct another QCA which squares exactly to our simplified 3F QCA; we call this the ``two semion QCA'' because it disentangles two copies of the $\{1,s\}$ Walker-Wang model.  We conjecture that it is a simpler, circuit-equivalent version of the square of the semion QCA constructed in \cite{Shirley}.  In \cref{app:trivialize}, we give an alternative proof that the 3F QCA trivializes with fundamental fermions, with a particularly simple circuit.  
%We show, however, that this circuit is not itself a ``trivial circuit", as defined later.  
In \cref{app:Chern} and \cref{app:CCZ} we give the details of the free fermion computations from \cref{backact}.

\section{Pumping $p+ip$: a $3$ dimensional generalization of the Thouless pump}
\label{backact}

In this section, we consider a finite time evolution generated by a free fermion quasi-local Hermitian operator that pumps a $p+ip$ state through a $3$ spatial dimensional system of fermions in a trivial product state on a lattice.  By quasi-local we mean that we will allow for exponentially decaying tails in the Hermitian generator of the evolution.  This is the continuous time version of a shallow depth quantum circuit of local unitaries.  For simplicity we will first consider a pump of a Chern insulator, which is easier to work with because it conserves particle number.  It is topologically equivalent to a pump of two copies of $p+ip$.  We work in reciprocal space, where a particle number conserving free fermion unitary can be described as a function from the Brilliouin zone to finite dimensional unitary matrices, $\veck \rightarrow U_\veck$.  We will construct a continuous path $U_\veck(t)$ from the identity ($U_\veck(0)=1$) to $U_\veck(1)=U_\veck$.  We can then recover the finite time evolution by writing down the time ordered exponential

\begin{align*}
U_\veck(t) = {\cal{T}} \exp\left(i\int_0^t d\tau \,K_\veck(\tau)\right)
\end{align*}
where

\begin{align*}
K_\veck(\tau) = -i U_\veck(\tau)^{-1} \partial_\tau U_\veck(\tau)
\end{align*}
is the Hermitian operator that generates the path.  

The main conclusion of this section is that the operator $U_\veck$ cannot be the identity, despite the fact that it describes the endpoint of the pumping process, when the Chern insulators have all been annihilated.\footnote{This is in contrast to the ordinary $1$ dimensional Thouless pump.}  Indeed, this conclusion must hold even if the pumping unitaries are not free-fermion, i.e. even in the interacting setting.  The argument is as follows.  Suppose an interacting evolution pumping $p+ip$ results in some many body operator $U$.  Truncate this evolution to a finite region of space.  This truncation of $U$ creates a Chern insulator on the boundary of this region, starting from a product state.  If $U$ were the identity, then the truncated operator would be identity in the bulk of the region.  In other words, we would have a quasi $2$-dimensional unitary operator that prepares a Chern insulator from a product state.  But this is a contradiction, because it would give us a commuting projector Hamiltonian for the Chern insulator, obtained by conjugating the trivial commuting projector Hamiltonian for the product state.  Such commuting projector Hamiltonians for chiral states are believed not to exist \cite{Kitaev_2005}.

We will show that, in the free fermion setting, the non-triviality of $U_\veck$ is encoded in a non-trivial winding number defined by $U_\veck$.  Specifically, $U_\veck$ will define a map from the Brilliouin zone $3$-torus to $SU(N_b)$, where $N_b$ is the number of bands.  Since $\pi_3(SU(N_b))=\Z$ for $n\geq 2$, this map is characterized by an integer winding number, or degree.  We will show that pumping a Chern insulator corresponds to a winding number of $2$, whereas pumping a $p+ip$ state corresponds to the minimal winding number of $1$.  We note that this winding number was identified in Ref. \cite{Rudner_pump}, though a physical interpretation was not given there.  This invariant also appears in a different physical context, namely that of the anomalous Floquet-Anderson insulator \cite{RudnerLevin}.

Our approach to demonstrating this correspondence is to construct a particular $U_\veck(t)$ that pumps a chiral state, and verify that the endpoint $U_\veck$ has the correct non-zero winding number.  Then we argue that any other such pumping process must be homotopic to it, and hence have the same winding number.  Because the map $\pi_3(SU(2))\rightarrow \pi_3(SU(N_b))$ induced by inclusion is an isomorphism for all $N_b \geq 2$, we will without loss of generality work with a two band model.  The explicit $U_\veck(t)$ that we construct will only be continuous in $\veck$ and $t$, but we expect that there is no obstruction to making it smooth, in both $\veck$ and $t$.  In that case, $K_\veck(\tau)$ will be quasi-local in real space with at most exponentially decaying tails.

\subsection{Nucleating a pair of $\pm 1$ Chern insulators} \label{ssec:pair}

Let us first discuss a single Chern insulator in a $2$ dimensional model with two bands, corresponding to a `flavor' index $\alpha=0,1$.  We let $X^\fl, Y^\fl$ and $Z^\fl$ be the Pauli matrices corresponding to this flavor index, with $Z^\fl = (-1)^\alpha$, and let $\vk=(k_x,k_y)$ be the $2d$ reciprocal wave-vector.  Consider the Hamiltonian:

\begin{align*}
H_{\text{C.I.}} = c^X(\vk) X^\fl + c^Y(\vk) Y^\fl + c^Z(\vk) Z^\fl
\end{align*}
In order for this to be a Hamiltonian for a Chern insulator with Chern number $\pm 1$ we choose the coefficients $c^X(\vk), c^Y(\vk), c^Z(\vk)$ such that $(c^X)^2+ (c^Y)^2+(c^Z)^2=1$ and such that the function $\vk \rightarrow (c^X,c^Y,c^Z)$, viewed as a map from $T^2$ to $S^2$, has winding number 1.  In \cref{app:Chern} we explicitly write down $c^X(\vk), c^Y(\vk), c^Z(\vk)$.

Now consider a stack of two such $2$ dimensional models.  Then there is an additional `layer' index $l=0,1$, so there are now $4$ bands total, again in $2$ spatial dimensions.  We can view the Hilbert space ${\mathbb{C}}^4$ over each $\vk$ as ${\mathbb{C}}^2 \otimes {\mathbb{C}}^2$, with the two tensor factors corresponding to the flavor and layer indices $\alpha$ and $l$ respectively.  In addition to the Pauli algebra generated by $X^\fl,Z^\fl$ we now also have the layer Pauli algebra generated by $X^\lay, Z^\lay$, with $Z^\lay = (-1)^l$.  Consider the Hamiltonian

\begin{align*}
H_{\text{pair}} = c^X(\vk) X^\fl + c^Y(\vk) Z^\lay Y^\fl + c^Z(\vk) Z^\fl
\end{align*}
Since $c^X(\vk) X^\fl \pm c^Y(\vk) Y^\fl + c^Z(\vk) Z^\fl$ is a Hamiltonian with Chern number $\pm 1$, $H_{\text{pair}}$ describes a Chern number $+1$ insulator on the layer $l=0$ stacked with a Chern number $-1$ Chern insulator on $l=1$.  There is no Chern number obstruction to trivializing the Hamiltonian $H_{\text{pair}}$, and indeed in \cref{app:Chern} we construct a unitary $\Un(\vk)$ with the property that

\begin{align*}
{\Un}^\dagger H_{\text{pair}} \Un = Z^\fl
\end{align*}
The unitary $\Un$ thus `nucleates' a pair of Chern number $\pm 1$ Chern insulators from the trivial product ground state of $Z^\fl$.  Because ${\Un}(\veck)$ is a map from $T^2$ (since it only depends on $\vk$) to $SU(4)$, and $\pi_1(SU(4)) = \pi_2(SU(4)) = 0$, it is homotopic to a constant (to retain exponential locality we really want it to be homotopic to a constant through a path of smooth maps, which we believe is possible but do not show explicitly).  Hence ${\Un}(\veck)$ is the endpoint of a quasi-local unitary evolution that starts at the identity.

\subsection{Pumping Chern insulators}

We are now ready to construct the unitary that pumps a Chern insulator through a trivial $3$ dimensional system of fermions.  Our model will be defined on a cubic lattice, with lattice constant set to $1$ for simplicity, initially with $2$ bands per unit cell.  These will again be labeled with a `flavor' index $\alpha=0,1$.  The ground state is defined by the trivial insulator Hamiltonian $H_{\text{triv}} = Z^\fl$.  Our unitary evolution will break translational symmetry in the $z$ direction down to two-fold translational symmetry, while retaining full translational symmetry in the $x$ and $y$ directions.  This is because our pumping process will nucleate Chern insulators on $2$ dimensional $xy$ layers, with the Chern number being $\pm 1$ according to the parity of the $z$ coordinate of the layer.  The Chern insulators will be nucleated in pairs having $z$ coordinates $(2n,2n+1)$, and will be annihilated with the complementary pairing (i.e. $(2n-1,2n)$).  On a system with boundary this results in Chern insulator states left over at the boundaries, showing that this process is indeed a Chern insulator pump.  The breaking of $z$ translational symmetry results in a doubled unit cell, and hence $4$ total bands, which can be labeled according to $(\alpha, l)$ where $l=0,1$ labels the position in the unit cell (i.e. is the parity of the $z$ coordinate) and will again be referred to as the `layer' index.

This setup can be viewed as an infinite stack of the bilayers of \cref{ssec:pair}.  We first nucleate identical pairs of $\pm 1$ Chern number Chern insulators on all the bilayers of the stack.  The corresponding unitary operator is just $\Un$, now viewed as a function of the $3$ dimensional reciprocal lattice vector $\veck = (k_x,k_y,k_z)$ but having no dependence on the $k_z$ coordinate.  We then annihilate complementary pairs of Chern insulators.  The operator $\Ua(\veck)$ which does this can be constructed as the conjugation by a translation by $1$ lattice constant in the $z$ direction, in the original non-doubled unit cell, of an operator that performs the annihilation on the same pairs of layers as the nucleation.  Viewed in terms of the doubled unit cell, the translation by $1$ is accomplished by swapping the layer index, and then translating one of the layers by $2$ (i.e. by a single doubled unit cell).  On the other hand, the operator that performs the annihilation can be taken to be the conjugate of $\Un$ by the operator $X^\lay$, which swaps the layers.

The end result of the nucleation and annihilation process is given by $U(\veck) = \Ua(\veck)\Un(\veck)$.  The same arguments as before show that $U(\veck)$ is the endpoint of a quasi-local unitary evolution.  In \cref{app:Chern} we explicitly construct $\Ua(\veck)$ and $U(\veck)$.  In particular, we show that $U(\veck)$ commutes with $Z^\fl$, and hence is block diagonal in flavor space:

\begin{equation} \label{eq:Ufinal}
U(\veck)=
\begin{pmatrix}
U_{\alpha=0}(\veck) & 0\\
0 & U_{\alpha=1}(\veck)
\end{pmatrix}
\end{equation}
with $U_{\alpha}(\veck)$ in $SU(2)$.  This is consistent with the fact that $U(\veck)$ should preserve the ground state of the trivial Hamiltonian $H_{\text{triv}} = Z^\fl$.  We also show in \cref{app:Chern} that the winding number, or degree, associated with $U_{\alpha}: T^3 \rightarrow SU(2) \sim S^3$, defined by \cite{Rudner_pump} as

\begin{align*}
\nu_3(U_{\alpha}) = \int \frac{d\veck}{24 \pi^2} \epsilon^{ijl} {\text{Tr}}\left[\left(U_\alpha(\veck)^{-1}\partial_{k_i}U_\alpha(\veck)\right) \left(U_\alpha(\veck)^{-1}\partial_{k_j}U_\alpha(\veck)\right) \left(U_\alpha(\veck)^{-1}\partial_{k_l}U_\alpha(\veck)\right) \right]
\end{align*}
is equal to $\nu_3(U_{\alpha}) = (-1)^\alpha$.  Hence pumping a Chern insulator does indeed correspond to a non-trivial unitary, with a non-trivial winding number in one of the blocks.  In the next subsection we perform a particle-hole transformation that shows that this process is equivalent to two identical copies of a process that pumps a $p+ip$ state, each of which having the same winding number $1$.

\subsection{Unitaries with non-trivial winding number pump $p+ip$ states}

Let us now perform a particle-hole conjugation $C$ only on the bands $\alpha=0, l=0,1$.  This is the anti-unitary operator that performs $a_{\alpha=0,l}(\veck)\leftrightarrow a^\dagger_{\alpha=0,l}(\veck)$, but does nothing to $a_{\alpha=1,l}(\veck), a^\dagger_{\alpha=1,l}(\veck)$, where $a_{\alpha,l}(\veck), a^\dagger_{\alpha,l}(\veck)$ are creation and annihilation operators on Fock space.  Conjugating $U(\veck)$ in eq. \ref{eq:Ufinal} by $C_{\alpha=0}$ we obtain:

\begin{equation} \label{eq:Uconj}
C U(\veck) C=
\begin{pmatrix}
U_{\alpha=0}(\veck) & 0\\
0 & U^*_{\alpha=1}(\veck)
\end{pmatrix}
\end{equation}
Note that it is the $\alpha=1$ bands whose unitary gets complex conjugated, since $C$ is acts like complex conjugation there, whereas the unitary for the $\alpha=0$ bands is fixed because if we write it as the exponential of $i$ times a Hermitian operator, the Hermitian operator is negated by particle-hole conjugation, as is the factor of $i$.  Eq. \ref{eq:Uconj} describes an $SU(4)$ unitary that is block diagonalized into two $SU(2)$ unitaries of the same winding number $+1$, since complex conjugation reverses the winding number.  Furthermore, this unitary $C U(\veck)C$ is the endpoint of a unitary evolution that pumps a Chern insulator in the background of the completely empty state.  Since the $SU(2)$ matrices in the two blocks have the same winding number, we can perform a homotopy from one to the other, which can be implemented by a unitary evolution in one of the blocks; the end result is that we can assume the two blocks are identical, $U_{\alpha=0}(\veck) = U^*_{\alpha=1}(\veck)$ for all $\veck$.  

Thus, $C U(\veck) C$ is the tensor product of two independent identical unitaries acting on the $\alpha=0$ and $\alpha=1$ bands respectively.  Let us now consider just one of these, say $U_{\alpha=0}(\veck)$, on its own.  Because it has a non-zero winding number, a quantized invariant of particle number conserving unitaries, we know that $U_{\alpha=0}(\veck)$ is not a finite time quasi-local unitary evolution.  However, if we allow all quadratic Hamiltonians, including pairing terms, then $U_{\alpha=0}(\veck)$ must be a finite time quasi-local unitary evolution.  This is because otherwise it would be a non-trivial free fermion QCA, which does not exist in this dimension and symmetry class ($d=3$, no symmetry other than fermion parity conservation)\cite{harper}.

The natural question then is, what does $U_{\alpha=0}(\veck)$ pump, when written as a finite time quasi-local unitary evolution with pairing?  Since two identical copies of it pump a state topologically equivalent to a Chern insulator, the only possibility is that one copy pumps a $p+ip$ state.  This result is easily generalized to show that a free fermion unitary with winding number $\nu$ pumps $\nu$ copies of a $p+ip$ state.

\subsection{Pumping a $p+ip$ state induces a CCZ gate}

In this subsection we turn to a general analysis of the unitary evolution that pumps a $p+ip$ state.  Let us put our $3+1$ dimensional lattice fermionic system, with a trivial product state Hamiltonian, on a spatial $3$-torus $T^3$, and gauge the fermion parity symmetry, i.e. perform minimal coupling to a $\Z_2^f$ gauge field.  Specifically, we can take $N_b$ fermions per site $\vecr$ on a cubic lattice, described by $2N_b$ Majorana modes $\gamma^{j}_\vecr, \bgamma^{j}_\vecr$, $j=1,\ldots, N_b$, as well as a $\Z_2$ gauge field on links between nearest neighbor sites $\vecr, \vecr+\hmu$, described by a qubit acted on by Pauli operators $X_{\vecr,\mu}, Z_{\vecr,\mu}$.  Here $\mu=1,2,3$ is a spatial index and $\hmu$ the unit vector in the $\mu$ direction.  The trivial Hamiltonian can then be taken to be
\begin{align} \label{H0}
H_0 = \sum_{\vecr} \sum_{j=1}^{N_b} i J \gamma^{j}_\vecr \bgamma^{j}_\vecr - J' \sum_{\vecr,\mu,\nu} F_{\vecr,\mu,\nu}
\end{align}
where $F_{\vecr,\mu,\nu} = Z_{\vecr,\mu} Z_{\vecr,\nu} Z_{\vecr+\hmu,\nu} Z_{\vecr+\hnu,\mu}$ is the $\Z_2^f$ gauge flux in the plaquette determined by $\vecr,\mu,\nu$.  The second term in the Hamiltonian above means that the $\Z_2^f$ gauge field has trivial dynamics.  There is also a Gauss law constraint
\begin{align*}
\prod_{j=1}^{N_b} \left(i \gamma^{j}_\vecr \bgamma^{j}_\vecr \right) \prod_{\mu} X_{\vecr,\mu} X_{\vecr-\mu,\mu} = 1
\end{align*}
The combined system of fermions and gauge fields, together with this Gauss law constraint, can be viewed as being built out of bosonic spin degrees of freedom \cite{Kitaev_2005}, as we will explicitly describe later.

We now consider a finite time unitary evolution that pumps a $p+ip$ state through the system.  This evolution can be coupled to the $\Z_2^f$ gauge field, and hence is some unitary operator in the combined system.  Specifically, if the evolution in the fermionic sector of the Hilbert space is generated by a quasi-local Hermitian operator $K$ (which will generally have a dependence on a parameter $\tau$ which we leave implicit), then $K$ can be minimally coupled to the $\Z_2^f$ gauge field in the standard way.  Namely, $K$ can be decomposed as a linear combination of local terms, each of the form
\begin{align*}
\gamma^{j_1}_{\vecr_1} \ldots \gamma^{j_k}_{\vecr_k} \bgamma^{j_{k+1}}_{\vecr_{k+1}} \ldots \bgamma^{j_{2n}}_{\vecr_{2n}}
\end{align*}
where $\{\vecr_1,\ldots,\vecr_{2n}\}$ lie in some region $D$ of bounded size.  Then let
\begin{align*}
P^D = \prod_{\vecr \in D} \left(1-F_{\vecr,\mu,\nu}\right)
\end{align*}
be the projector onto flat gauge field configurations (i.e. no fluxes) in the region $D$, and let $S$ be a set of links in $D$ that pairwise connect all the $\{ \vecr_1,\ldots,\vecr_{2n}\}$ (i.e. $S$ is a $\Z_2$-valued $1$-chain with $\partial S = \{ \vecr_1,\ldots,\vecr_{2n}\}$).  We then define the corresponding gauged term in the Hamiltonian as
\begin{align*}
P^D \cdot \left(\gamma^{j_1}_{\vecr_1} \ldots \gamma^{j_k}_{\vecr_k} \bgamma^{j_{k+1}}_{\vecr_{k+1}} \ldots \bgamma^{j_{2n}}_{\vecr_{2n}}\right) \cdot \left( \prod_{(\vecr,\mu)\in S} Z_{\vecr,\mu}\right) \cdot P^D
\end{align*}
Because we project onto flat gauge field configurations in $D$, this definition is independent of the choice of $S$, and just amounts to the standard minimal coupling procedure.  Note that the $\Z_2^f$ gauge field is still non-dynamical during this pumping process, i.e. all terms in the Hamiltonian commute with $F_{\vecr,\mu,\nu}$.  This allows us to consider the evolution independently in each gauge flux sector.

Now, the combined system of fermions and gauge fluxes described by $H_0$, viewed as a spin system, is topologically ordered, with a ground state degeneracy of $2^3=8$.  A basis for the ground state subspace can be taken by picking $\pm 1$ $\Z_2^f$ gauge field holonomies in the $3$ directions of the torus.  Here the holonomy around a particular cycle is defined as the product of $Z_{\vecr,\mu}$ around that cycle.  We claim that, in this basis, up to a non-universal overall phase, the gauged evolution implements a $CCZ$ gate, in the thermodynamic limit of large system size.  Specifically, if all three of the holonomies are $-1$, then there is an additional $\pi$ phase shift compared to all the other cases, which all have the same phase as each other.  To say it differently, the system encodes three logical qubits, and the pump induces a CCZ gate on these qubits, in a basis in which the $Z$ state of each logical qubit gives the corresponding holonomy for one of three directions.

%We consider a $3+1$ dimensional system of fermions coupled to a $\mz_2$ gauge field. Suppose the ground state has vanishing gauge flu and some gapped state (topologically ordered or trivial) of the fermions.  Now consider pumping a $p+ip$ superconductor through this system.  The gauge flux is still vanishing after the pump, and the holonomies (if the ambient space has nontrivial $\pi_1$) are unchanged.  However, the pump induces some phase; this phase has some non-universal term independent of the gauge field, and also has some universal part depending on the holonomy.  We can think of this as some ``back action" on the gauge field.  This is what we wish to consider.

%In the case of a three torus, we can describe the gauge field by three holonomies, each $\pm1$, depending on winding around each of three directions of the torus.  We claim that if all three of these equal $-1$, then there is an additional $\pi$ phase shift compared to all the other cases, which all have the same phase as each other.

We can see this as follows.  First, consider a Kitaev chain with periodic boundary conditions.  This state has the property that the fermion parity of its ground state changes upon imposing $-1$ holonomy.  Now consider a $p+ip$ state on a cylinder.  Inserting a $\pi$ flux through the cylinder - i.e. imposing $-1$ holonomy around the cylinder - results in a Majorana zero mode at the ends of the cylinder, turning this dimensionally reduced system into a Kitaev chain.  Thus the fermionic parity of a $p+ip$ state on a $2$-torus changes when $-1$ holonomy is inserted along both directions of the torus.

Now let us return to our three dimensional system.  Let us take a $T^3$ whose dimensions are large in lattice units.  We will discuss the gauged evolution in a continuum limit, so let us parametrize the three directions along the $T^3$ by continuous coordinates $x,y,z$, each periodic modulo $1$.  One way to view the pumping process is as follows.  First, the process creates a small bubble of $p+ip$ superconductor near some point, say $(1/2,1/2,1/2)$.  The bubble then expands in the $x$ and $y$ directions until it gives two planes of $p+ip$ superconductor, one at $z=1/2+\epsilon$ and one at $z=1/2-\epsilon$ for some small $\epsilon$ (but still large compared to the lattice scale).  The planes are oppositely oriented; equivalently, one may say they are oriented the same way but one plane is $p+ip$ and one plane is $p-ip$.  Now, if the holonomies in both $x$ and $y$ directions are equal to $-1$, then each plane has odd fermion parity, by the discussion above; otherwise they have even fermion parity.  Finally, move the upper plane further upwards, increasing its $z$ coordinate, until it annihilates against the lower plane from below.  An additional $\pi$ phase shift is induced if the plane has odd fermion parity and if the $z$ holonomy is $-1$, since this is then effectively a $\Z_2^f$ charge being taken around a cycle of non-trivial holonomy.  Hence, the claim follows.

We can also understand this from a continuum field theory point of view, as follows.  In $d$ spatial dimensions (here, $d=3$), we can pump some system $S$ with $d-1$ spatial dimensions.  The pump then sweeps out some $d$ dimensional volume.  Since the system $S$ has $d-1$ spatial dimensions, its Euclidean path integral is $d$ dimensional,
and \emph{the phase induced is given by this Euclidean path integral in the background of the given gauge field}.  This is a general principle valid for pumping other theories coupled to a gauge field.

To compute this phase, we need the effective action for a $p+ip$ superconductor coupled to a gauge field.  Now, because a $p+ip$ superconductor is a chiral invertible state and thus has a non-trivial thermal Hall response, its corresponding effective action contains geometric terms that depend on the spacetime curvature \cite{Kapustin}.  To avoid these, it is convenient to introduce a second state, namely a $p-ip$ superconductor, that cancels this thermal Hall response but is not coupled to the gauge field.  That is, on each bubble we have two decoupled systems, one $p+ip$ and one $p-ip$.  There are thus two individual $\Z_2$ symmetries that are preserved.  We introduce a gauge field for only one of these, namely the one associated to $p+ip$.  Note that the diagonal combination is the overall fermion parity, which is not gauged, so 
we should view this as a pump of a $2d$ fermionic symmetry protected topological (SPT) phase of a unitary $\Z_2$ symmetry; such SPT phases have a $\Z_8$ classification\cite{GuLevin}.  The effective action in this case depends on the Rokhlin invariant, a mod $16$ invariant on three-manifolds with spin structure \cite{KapustinPC}.  With background spin structure $\alpha$, and gauge field $x$, the effective action gives a phase
$\pi(R(\alpha+x)-R(\alpha)/8)$ \cite{brum, turaev1984}.  This difference of Rokhlin invariants is $0$ mod $8$ \cite{dahl2002dependence}, so the effective action gives a phase $0$ or $\pi$.  In the case of a three torus, the difference of Rokhlin invariants gives the CCZ gate. %, but this gives a general way to compute it .
 
The fact that pumping a $p+ip$ state gives a CCZ gate on a three torus can also be verified by an explicit computation, using a free fermion form of the unitary circuit that pumps $p+ip$.  This free fermion unitary circuit can be made translationally invariant, so the computation ends up taking place in some (particle-hole symmetric) band structure over a Brilliouin zone.  This computation is done in \cref{app:CCZ}.

\section{Pumping With Gauge Flux, Fermions in Ground State}
\label{multicopypump}
We now consider the case of pumping $k$ copies of $p+ip$, for some integer $k$, in a state with non-vanishing gauge flux, but with the fermions in the ground state of the trivial product state Hamiltonian $H_0$ (see eq. \ref{H0}).  This just means that the fermions are in the zero occupation state.  We consider $k=1,2,4,8$, but our results for these respective values of $k$ will apply to others that are equal to them modulo $2k$.

One difficulty in analyzing such pumps is the inherent ambiguities in defining them.  For example, two pumps may differ by a local quantum circuit which does not itself pump any chirality.  Defining a pump as an adiabatic evolution of some Hamiltonian still leaves open the precise choice of adiabatic path.  However, we will argue that regardless of how these choices are made, for $k=1,2$, the pump process will inevitably produce some excitations in the fermionic sector of the Hilbert space at its endpoint.  Furthermore, we will argue that for $k=4,8$ it is possible to define the pump in some way so that the fermionic state is not excited if it starts in the zero fermion occupation state.

For $k=1$ we have a single $p+ip$ state.  A vortex in a $p+ip$ state binds a Majorana zero mode.  To get some intuition for the implications of this fact, consider first a simple geometry where there is a circular gauge flux, with the circle sitting in the $xz$ plane.  If we orient the $p+ip$ state parallel to the $xy$ plane and pump it in the $z$ direction, the corresponding braiding history describes creating a pair of Majorana zero modes and then annihilating them.  In this case the overall fermion parity of these two Majorana modes is clearly conserved, and there is no obstruction to having all the fermions remaining in the zero fermion occupation state at the end of the process.  However, consider now a more complicated gauge flux geometry, consisting of two linked loops.  The corresponding braiding history will then describe creating two pairs of Majorana modes, braiding one member of one pair with another member of the second pair, and then annihilating the pairs.  In this case, the overall fermionic parity is preserved, as it must be, but we may create two fermions, one on each loop, when we do the annihilation.  Note that it is possible to unambiguously define the parity near a loop, so long as the fermions away from the loops stay in their ground state.  This possible creation of fermions will be a general feature of any pump for $k=1$.

For $k=2$, the flux lines describe the braiding history of anyons in $U(1)_4$ Chern-Simons theory \cite{Kitaev_2005}.  This is an abelian theory with $\mathbb{Z}_4$ fusion rules.  Label the anyons by $0,1,2,3$, with the vortices corresponding to either $1$ or $3$, $0$ being the identity particle, and $2$ being the fermion.  Even though the theory is abelian, we argue that it is still impossible, in general, to define the pump while remaining in the zero fermion occupation state.  This is because whether a gauge flux loop represents a $1$ or a $3$ particle depends on a choice of orientation along the loop, since these two anyons are each others' anti-particles.  However, it is impossible to use only local information to give a consistent orientation to an a priori unoriented loop.  Hence, we will inevitably have flux configurations which describe fusing two $1$ particles into a $2$ particle; the $2$ particle is the emergent fermion.

For $k=4$, we $U(1)_2 \times U(1)_2$ Chern-Simons theory \cite{Kitaev_2005}, which has $\mathbb{Z}_2 \times \mathbb{Z}_2$ fusion rules.  Namely, the anyons can be parametrized as $\{1,s_1\} \times \{1,s_2\}$ where $s_1, s_2$ are semions and $f = s_1 s_2$ is the emergent fermion.  The flux lines describe the braiding history of either $s_1$ or $s_2 = s_1 f$ particles.  If we choose them all to be $s_1$, which we can do locally, since both $s_1$ and $s_2$ are their own anti-particles, then this keeps the fermions in their zero occupation state.

Finally, for $k=8$, the theory again has $\mathbb{Z}_2 \times \mathbb{Z}_2$ fusion rules, and is the $SO(8)_1$ Chern-Simons theory, also known as the three fermion theory \cite{Kitaev_2005}.  The fusion group of the anyons is $\{1,e\} \times \{1,m\}$ where $e,m$ are both fermions.  The flux lines describe the braiding history of $m$ particles, while $f=em$ is the emergent fermion.

\section{Simplified 3-fermion QCA}
\label{simpleqca}

We now turn to non-trivial QCA, which are locality preserving unitaries which are 
not local unitary evolutions.  We will construct QCA which have the same action on the $\Z_2^f$ gauge field fluxes as the $k=4$ and $k=8$ local unitary evolutions discussed in \cref{multicopypump}, but which act trivially on the fermions.  This is in contrast to the local unitary evolutions discussed above, which have a non-trivial winding number characterizing their action on the fermion degrees of freedom.
 
\subsection{Review of $\{1,f\}$ Walker-Wang model}
\label{review1f}
\begin{figure}[tb]
\begin{center}
\includegraphics[width=0.5\linewidth]{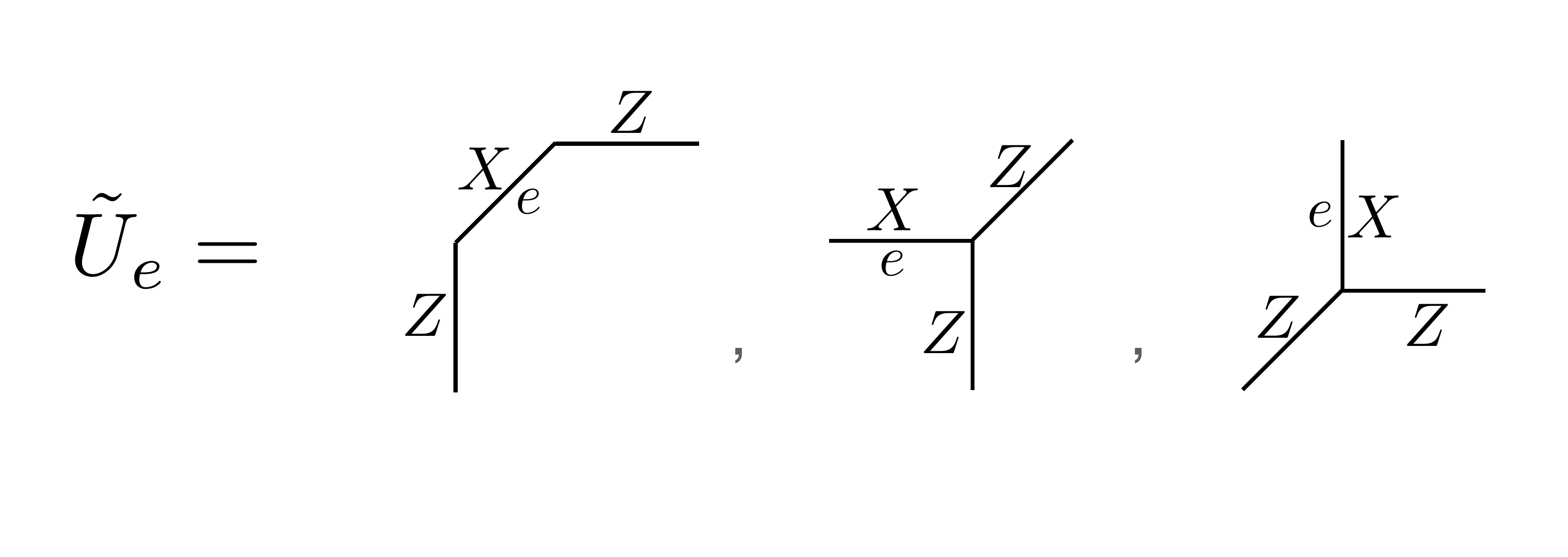}
%\vspace{-.1in}
\end{center}
\caption{Definition of fermion string operators $\tU_e$.}
\label{fig:fig1p}
\end{figure}

In this section we again consider fermions in $3$ spatial dimensions, but this time we couple them to a dynamical $\Z_2^f$ (fermion parity) gauge field.  The resulting Hilbert space is bosonic, and may be explicitly described as the Hilbert space of the Walker-Wang model based on the pre-modular category $\{1,f\}$, where $f$ is a fermion.  Concretely, this is the Hilbert space of a cubic lattice model, with periodic boundary conditions in $x,y,z$, with one qubit per edge $e$, acted on by the Pauli algebra generated by $X_e, Z_e$.  The Hamiltonian of the $\{1,f\}$ Walker-Wang model is given by $H_{\{1,f\}\,\text{WW}} = -\sum_v A_v - \sum_p B_p$, where $A_v$ is the vertex term $A_v \equiv \prod_{e\sim v} Z_e$ and $B_p$ the plaquette term $B_p \equiv \prod_{e\in \partial p}  \tU_e$.  Here $\tU_e$ are short fermionic string operators and are defined in figure \ref{fig:fig1p}.  We also define fermionic string operators associated to a path of edges by $\tU_{\text{path}} \equiv \prod_{e \in \text{path}} \tU_e$, with the sign convention that all of the Pauli $Z$ operators act before any of the Pauli $X$'s.  This notation and convention is identical to that of \cite{Shirley}; we discuss the relation between our QCA and that of \cite{Shirley} below.  We note that one can view the Hilbert space of this spin model as the Hilbert space of a fermion coupled to a $\Z_2^f$ gauge field, and under this interpretation the short string operators are precisely fermionic hopping operators.  We will not need the explicit dictionary between the fermionic and bosonic operators in this work, but we encourage the interested reader to consult reference \cite{Kapustin3d} for more details.

We will not work with the Hamiltonian of the $\{1,f\}$ Walker-Wang model directly.  Rather, its purpose is to provide a convenient orthonormal basis of states for the Hilbert space.  
This basis is parametrized by specifying the locations of the fermion excitations and the magnetic fluxes.  Specifically, we consider the vertex terms $A_v$ and the plaquette terms $B_p$, together with three homologically non-trivial `holonomy-detecting' string operators $H_i \equiv \prod_{e \in \text{path } i} \tU_e$, where $\text{path } i$ is a fixed, homologically non-trivial path in the direction $i=x,y,z$.  As is known from the study of the $\{1,f\}$ Walker-Wang model, or from the bosonization duality of \cite{Kapustin3d}, specifying the eigenvalues $\pm 1$ of the operators $\{A_v, B_p, H_i\}$ uniquely determines a state in Hilbert space, up to phase.  The eigenvalues of $B_p$ are constrained so that the $-1$ eigenvalues form closed loops of plaquettes; the eigenvalues of $A_v$ and $H_i$ are un-constrained.  Informally, we may say that we can specify a state uniquely by specifying the locations of the fermion excitations, the magnetic flux lines, and the non-trivial holonomies.

We will also need the following property of the $\{1,f\}$ Walker-Wang Hamiltonian.  Let $|\psi_0\rangle$ be the ground state of $H_{\{1,f\}\,\text{WW}}$ with trivial holonomies in all three directions.  Let us view this ground state in the basis that diagonalizes $\{Z_e\}$.  This is the electric flux line basis, and it is known that $|\psi_0\rangle$ is, up to normalization, a superposition of all closed, homologically trivial electric flux loop configurations $C$, weighted by a sign $(-1)^{\text{framing}(C)}$.  Here the framing of a loop configuration $C$ is defined as follows.  We translate $C$ by $\left(-\frac{1}{2},-\frac{1}{2},-\frac{1}{2}\right)$ to obtain a copy of $C$, denoted $C'$, on the dual lattice.  We then compute the modulo $2$ linking number of $C$ and $C'$, by counting the parity of the number of times $C$ crosses $C'$ when the whole picture is projected on a plane.  

This $(-1)^{\text{framing}(C)}$ amplitude is just a reflection of the fact that, in the $\{1,f\}$ Walker-Wang model, the electric flux lines are to be viewed as $2+1$ dimensional worldlines of the fermion $f$.  Note that based on our choice of $|\psi_0\rangle$, $C$ is homologically trivial.  However, the above definition of $(-1)^{\text{framing}(C)}$ is well defined even for homologically non-trivial $C$.  One consequence of the above fact is the following.  Let $C$ be a closed loop configuration.  Then 

\begin{align} \label{eq:psi0}
\tU_C|\psi_0\rangle = (-1)^{\text{framing}(C)} |\psi_0\rangle.
\end{align}
More generally, this is true for any state with a flat gauge field and trivial holonomy (i.e. no magnetic fluxes).  This is because this space of states is dual to fermions by the bosonization duality of \cite{Kapustin3d}, and such states can be built up by acting with fermionic creation operators on the ground state; the claim can then be verified by the Majorana commutation relations.

For an arbitrary eigenstate $|\psi\rangle$ of the vertex, plaquette, and holonomy operators, we have the more general formula:

\begin{align} \label{eq:psi}
\tU_C|\psi\rangle = (-1)^{\text{link}(C,L)}(-1)^{\text{framing}(C)} |\psi\rangle.
\end{align}
where $L$ is the magnetic flux line configuration associated to $|\psi\rangle$.  The braiding phase $(-1)^{\text{link}(C,L)}$ in fact also depends on the holonomy eigenvalues.  An easy way to see this is to note that the information in $L$ and the holonomy eigenvalues associated to $3$ specific paths generating $\pi_1(T^3)$ can alternatively be packaged as an assignment of a homology class of $2d$ membranes $M$ with $\partial M = L$.  We then have

\begin{align*}
(-1)^{\text{link}(C,L)} = (-1)^{\text{int}(C,M)}
\end{align*}
For conciseness we will however stay with the notation $(-1)^{\text{link}(C,L)}$ for the braiding phase.

\subsection{Definition of locality preserving unitary $\Ufr$}
\label{localitypreservingsubsec}

We now define a locality preserving unitary $\Ufr$, which we later show is stably equivalent to the non-trivial $3$-fermion QCA.  $\Ufr$ is diagonal in the basis of eigenvectors of $A_v,B_p, H_i$ discussed above, with eigenvalue $\pm 1$.  Specifically, the eigenvalue is defined to be $(-1)^{\text{framing}(L)}$, where $L$ is the loop configuration of {\emph{magnetic}} fluxes defined by the $-1$ eigenvalues of $B_p$.  Note that $L$ lives on the dual lattice, but we may still use the definition of framing above, just shifted by $\left(\frac{1}{2},\frac{1}{2},\frac{1}{2}\right)$.

There are several ways to see that $\Ufr$, so defined, is locality preserving. 
Let us consider first a direct argument.  Let $O$ be some operator supported on some ball $B$.
We describe the magnetic flux $L$ by two pieces of data, the flux within $O(1)$ distance from $B$, and the flux further away; these two pieces of data are subject to a constraint of course as the flux lines are closed.  The operator $O$ commutes with the second piece of data, but may change the first piece.
Crucially, while the framing of the magnetic flux $L$ cannot be computed just from the first piece of data, the {\emph change} in the framing can be and so indeed $\Ufr$ is locality preserving.

  An indirect way to see this is from the fact that upon introducing ancilla fermions, $\Ufr$ can be written as a shallow depth circuit of local unitaries that tunnel these ancilla fermions along the magnetic flux loops, picking up a sign of $(-1)^{\text{framing}(L)}$.  More precisely, this can be done by introducing a second copy of the $\{1,f\}$ Walker-Wang Hilbert space on the dual lattice, and defining a shallow depth circuit $U'$ on the tensor product of these two by acting with $\tU_L^{\text{anc}}$ in the ancilla copy conditioned by the magnetic flux configuration $L$ in the original copy:

\begin{align*}
U' = \sum_L P_L \otimes \tU_L^{\text{anc}}
\end{align*}
Here $P_L$ projects onto the subspace of states with magnetic flux loops along $L$, and $\tU_L^{\text{anc}}$ is the closed string operator of ancilla fermions.  Note that this definition makes sense since the magnetic loops $L$ and the ancilla fermion tunneling operators both live on the dual lattice.

A standard argument shows that $U'$ is indeed a shallow depth circuit, as follows.  We pick an $n$-coloring of the edges of the dual lattice such that no two edges sharing a vertex are the same color for some finite $n$.\footnote{One way to do this is to consider the coordinates $(x,y,z)$ of the center of each edge, which are all in $\Z/2$, and assign a different color to $(x \mod 2, y \mod 2, z \mod 2)$. This requires at most $n=4^3=64$ colors, although more efficient schemes are certainly possible.}  We then build up $\tU_L$ by acting sequentially with the product of $\tU_e$ for edges $e$ in $L$ with color $i$, for $i=1,\ldots n$.  To ensure that all the $Z$'s act before the $X$'s, we insert extra signs when this is not the case: these signs are associated to endpoints of the edges and are locally determined from the coloring of all the other edges ending at those endpoints, as well as the configuration $L$.  Namely, when acting with $\tU_e$ for $e$ with color $i$, we act with an additional factor of $-1$ for each edge $e'$ adjoining $e$ at either vertex satisfying the conditions 1) the color of $e'$ is $i'<i$ and 2) there is a Pauli $Z$ in $\tU_e$ acting on edge $e'$.

Now let us see why $U'$ being a shallow depth circuit implies that $\Ufr$ is locality preserving.  
For any state $|\psi^\text{anc}\rangle$ of the ancilla Walker-Wang model in the no magnetic flux and trivial holonomy subspace, we have by \ref{eq:psi0} $U'|\chi\rangle |\psi^\text{anc}\rangle = (-1)^{\text{framing}(L(|\chi\rangle))}|\chi\rangle |\psi^\text{anc}\rangle$, so that $U'|\chi\rangle |\psi^\text{anc}\rangle = \Ufr |\chi\rangle |\psi^\text{anc}\rangle$ for all $|\chi\rangle$.  
This means that for any two local well separated operators $A$ and $B$ in the original system, $\left(\Ufr A \Ufr^{-1}\right) |\psi^\text{anc}\rangle\langle \psi^\text{anc}|$ commutes with $B |\psi^\text{anc}\rangle\langle \psi^\text{anc}|$, which implies that $\Ufr A \Ufr^{-1}$ commutes with $B$, which in turn implies that $\Ufr$ is locality preserving.

Constraining the ancilla Walker-Wang model to this subspace with trivial holonomy and no magnetic fluxes is equivalent to tensoring with fundamental fermions.  So, the statement that $U'$ is a shallow depth circuit is equivalent to saying that $\Ufr$ {\emph{trivializes}} in the presence of fundamental fermions.
In the next subsection, we show that $\Ufr$ is circuit equivalent to the $3$ fermion QCA, implying that that QCA trivializes in the presence of fundamental fermions as well, meaning that upon tensoring the Hilbert space with additional ancilla fermions, the product of that QCA with the identity on those additional fermions is equivalent to conjugation by a circuit.  We give an alternative, more direct, proof of this in \cref{app:trivialize}.

\subsection{$\Ufr$ is circuit-equivalent to the non-trivial $3$ fermion QCA}

We will now show that the locality preserving unitary $\Ufr$ is circuit-equivalent to the non-trivial $3$ fermion QCA.  We will actually show this for the operator

\begin{align*}
U_\text{link} \equiv U' \left(\Ufr \otimes 1^\text{anc}\right),
\end{align*}
which is in the same QCA equivalence class as $\Ufr$.  Note that now we do {\emph{not}} constrain the ancilla Walker-Wang model to be in the trivial holonomy subspace.

The operator $U_\text{link}$ has a simple interpretation in the electric charge and magnetic loop basis.  Namely, when acting on a state $|\chi\rangle |\chi^\text{anc}\rangle$, with $|\chi\rangle$ and $|\chi^\text{anc}\rangle$ having magnetic loop configurations $L$ and $L^\text{anc}$ respectively, it gives a factor of $(-1)^{\text{link}(L,L^\text{anc})}$.  This arises simply from the fact that $U'$ tunnels ancilla fermions along $L$, and these pick up the usual Berry phase from braiding with ancilla magnetic fluxes $L^\text{anc}$, in addition to the factor of $(-1)^{\text{framing}(L)}$ due to the fermionic nature of the ancilla gauge charges; see eq. \ref{eq:psi}.  The additional factor of $(-1)^{\text{framing}(L)}$ in the definition of $\Ufr$ then results in $U_\text{link}$ acting as $(-1)^{\text{link}(L,L^\text{anc})}$.

Thus we just have to show that $U_\text{link}$ is circuit-equivalent to the 3-fermion QCA.  To do this, we show that, up to a shallow depth circuit, this operator is the same as the 3-fermion locality preserving unitary $\talpha_{3F}$ constructed in \cite{Shirley}.  $\talpha_{3F}$ is defined through its action on Pauli $X_i$ and $Z_i$ operators in two copies $i=1,2$ of the $\{1,f\}$ Walker-Wang Hilbert space, illustrated in figures 7 and 8 of Ref. \cite{Shirley}.  In our notation $i=1,2$ correspond to the original system and the ancilla respectively.  We will refer to an operator acting on system $i$ as an ``operator of type $i$", and similarly we will refer to ``charge of type $i$" or ``flux of type $i$" if they correspond to copy $i$.

%\begin{figure}[tb]
%\begin{center}
%\includegraphics[width=1.0\linewidth]{QCA_definition_fig1.pdf}
%\vspace{-.1in}
%\end{center}
%\caption{Definition of $\talpha_{3F}$ taken from Ref. \cite{Shirley}.  This shows the stabilizers of the associated separator, i.e. the images of the Pauli $Z$ operators under $\talpha_{3F}$.  The sign convention for multiplying the string operators is that $Z$'s act before $X$'s.  The subscripts $1$ and $2$ refer to the original system and the ancilla system respectively.}
%\label{fig:QCAdef1}
%\end{figure}

%\begin{figure}[tb]
%\begin{center}
%\includegraphics[width=1.0\linewidth]{QCA_definition_fig2.pdf}
%\vspace{-.1in}
%\end{center}
%\caption{Definition of $\talpha_{3F}$ taken from Ref. \cite{Shirley}.  This shows the flippers of the associated separator, i.e. the images of the Pauli $X$ operators under $\talpha_{3F}$.  The sign convention for multiplying the string operators is that $Z$'s act before $X$'s.  The subscripts $1$ and $2$ refer to the original system and the ancilla system respectively.}
%\label{fig:QCAdef2}
%\end{figure}

We first note that the orthonormal basis defined by the location of fermions and magnetic fluxes of both types, as well as their holonomies, is an eigenbasis of $\talpha_{3F}$.  To see this it is sufficient to show that the vertex, plaquette, and long fermionic string operators are all preserved by $\talpha_{3F}$.  This follows by inspection of figures 7 and 8 of Ref. \cite{Shirley}; for the plaquette term this is also explicitly verified in \cite{Shirley}.  In particular this means that $\talpha_{3F}$ acts as the identity on the ground state subspace.

All that is left is computing the eigenvalue $\pm 1$ of $\talpha_{3F}$ on such a basis state.  Now, any such state can be built from a ground state of the two $\{1,f\}$ WW models by applying open membrane operators to create the flux loops and string operators to create the fermionic point charges.  So we just need to conjugate all these operators by $\talpha_{3F}$.  We build up the state step by step, starting with the flux excitations.  Now, one thing we can explicitly verify using figures 7 and 8 of Ref. \cite{Shirley} is that conjugating a membrane operator of type $1$ gives that same membrane operator times a string operator that tunnels a charge of type $2$ along its boundary, and vice versa.  Thus, every time we put in a flux, we pick up a sign if there is an odd linking number of that flux with the fluxes of the other type that are already there, and an additional sign corresponding to the framing of the flux.  All together, we get a sign corresponding to the mod $2$ linking number of the configuration of type $1$ and $2$ fluxes, and signs corresponding to the framing of type $1$ and type $2$ fluxes.  However, by multiplying by the operator $U'$ and its partner under the exchange of the original system and the ancilla, we can get rid of the framing signs, leaving us with just the linking sign.

Now let us deal with the charges, i.e. violations of the vertex term.  Any such charge configuration can be created by acting with open string operators.  Conjugating such an open string operator by $\talpha_{3F}$ results in that same string operator multiplied by some closed loop operator decorations of the other type near the endpoints.  Since closed string operators are invariant under $\talpha_{3F}$, these decorations depend only on the endpoint vertex and nothing else.  Thus, the contribution of the charges to the eigenvalue is equal to the product, over all occupied vertices, of these decorations acting on the magnetic loop configuration.  We do not need to work out this contribution in detail since it is manifestly a shallow depth circuit.

We have thus shown that $\Ufr$, $U_{\text{link}}$, and $\talpha_{3F}$ are all circuit-equivalent.

\subsection{Action of $U_{\text{link}}$ and $\Ufr$ on the trivial product state}\label{Uact}

It is interesting to also directly examine the action of $U_{\text{link}}$ and $\Ufr$ on the trivial product state.  In this section we will use the subscripts $i=1,2$ to refer to the original and ancilla systems respectively.  The trivial product state $|\psi_{\text{prod}}\rangle$ is then the one with $Z_{i,e}=1$, for all edges $e$ and $i=1,2$.  We will directly check that $U_{\text{link}}$ produces the ground state of the $3$-fermion Walker-Wang model in the electric line basis, and we will also make some comments about the action of $\Ufr$ on the trivial product state.

The trivial product state $|\psi_{\text{prod}}\rangle$, corresponding to all electric fluxes being confined, is an equal amplitude superposition of all magnetic flux loop configurations:

\begin{align}
|\psi_{\text{prod}}\rangle = C \prod_{i,e} \left(\frac{1+Z_{i,e}}{2}\right)|\psi_{0,0}\rangle
\end{align}
up to some normalization constant $C$, where $|\psi_{0,0}\rangle$ is the tensor product of the Walker Wang ground states in $i=1,2$, say in the trivial holonomy sector.  Expanding out the above product, each term corresponds to some collection of edges for copy $1$ and for copy $2$.  It will be useful to work in the dual picture, where we have a collection of dual plaquettes for both copies.  These dual plaquettes form some membrane $M_i$ with the two magnetic flux configurations corresponding to the $\partial M_i$.  Since by Gauss's law $|\psi_{0,0}\rangle$ is a $+1$ eigenvalue eigenstate of all such terms corresponding to closed, homologically trivial $M_i$, we may re-write the above equation as

\begin{align}
|\psi_{\text{prod}}\rangle = C' \sum_{\{M_i\}} Z_{M_1} Z_{M_2} |\psi_{0,0}\rangle
\end{align}
where $C'$ is some other constant and $Z_{M_i}$ is shorthand for the product of all Pauli $Z$ operators making up the membrane $M_i$.  The sum is taken over all homology classes of $M_i$ (the equivalence relation is that $M_i$ is equivalent to $M'_i$ if their union is closed and homologically trivial).  The fact that we are summing over homology classes of membranes, rather than membranes themselves, is consistent with the fact that we are not directly accessing the $\Z_2$ vector potential as a local degree of freedom.  Thus

\begin{align} \label{eq:Ustate}
U_{\text{link}} |\psi_{\text{prod}}\rangle = C' \sum_{\{M_i\}} (-1)^{\text{int}(M_1,\partial M_2)} Z_{M_1} Z_{M_2} |\psi_{0,0}\rangle
\end{align}
We now want to express this state in the electric line basis.  Note that $|\psi_{0,0}\rangle$ is a superposition of all closed electric loop configurations $E_1,E_2$ in both copies, weighted by the product of their framing signs.  Thus, a state of the form $Z_{M_1} Z_{M_2} |\psi_{0,0}\rangle$ is also a superposition of all closed electric loop configurations, but in addition to the framing signs there is also a factor of $(-1)^{\text{int}(M_1,E_1)} (-1)^{\text{int}(M_2,E_2)}$.  Inserting this into eq. \ref{eq:Ustate} we obtain:

\begin{align} \label{eq:Usum}
\langle \{v_{i,e}\} |U_{\text{link}} |\psi_{\text{prod}}\rangle = C'' \sum_{w_{i,e}} (-1)^{\text{int}(w_{1}, \partial w_{2}) + \text{int}(w_{1}, v_{1}) + \text{int}(w_{2}, v_{2}) + \text{framing}(v_1) + \text{framing}(v_2)}
\end{align}
Here $\{v_{i,e} = 0,1\}$ define the closed electric loop configurations $E_i$: $v_{i,e}=1$ corresponds to $Z_{i,e}=-1$.  The sum on $\{w_{i,e} = 0,1\}$ is a sum over all homology classes of membranes (with boundary) with the equivalence relation described above; for each such equivalence class, we pick a specific membrane $M$ represented by $w_{i,e}=1$ for all the edges defining $M$.  Alternatively, we may simply sum over all possible $\{w_{i,e} = 0,1\}$, since this just introduces an extra overall multiplicative factor equal to the number of membrane configurations in each equivalence class (note that this is independent of the equivalence class).  For homologically non-trivial $v_{i,e}$ we immediately see that the sum in eq. \ref{eq:Usum} is $0$; for homologically trivial $v_{i,e}$ we can perform the sum by noting that the intersection is a $\Z_2$ quadratic form, and completing the square.  Specifically, shifting $w_{i,e}\rightarrow w_{i,e}+y_{i,e}$, we get

\begin{align*}
\text{int}(w_{1},\partial w_{2}) &\rightarrow \text{int}(w_{1},\partial w_{2}) + \text{int}(y_1, \partial w_{2}) + \text{int} (w_1, \partial y_2) + \text{int} (y_1, \partial y_2) \\ &= \text{int}(w_{1},\partial w_{2}) + \text{int}(w_2, \partial y_{1}) + \text{int} (w_1, \partial y_2) + \text{int} (y_1, \partial y_2)
\end{align*}
so choosing $y_i$ such that $\partial y_{i,e} = v_{i,e}$ we see that the terms linear in $w_i$ in the exponent of eq. \ref{eq:Usum} are eliminated, and we obtain

\begin{align}
\langle \{v_{i,e}\} |U_{\text{link}} |\psi_{\text{prod}}\rangle &= C'' \sum_{w_{i,e}} (-1)^{\text{int}(w_{1}, \partial w_{2}) + \text{int} (y_1, \partial y_2)+ \text{framing}(v_1) + \text{framing}(v_2)} \\ &= C'' \sum_{w_{i,e}} (-1)^{\text{int}(w_{1}, \partial w_{2}) + \text{link} (v_1, \partial v_2)+ \text{framing}(v_1) + \text{framing}(v_2)} \\ &= C''' (-1)^{\text{link} (v_1, \partial v_2)+ \text{framing}(v_1) + \text{framing}(v_2)}
\end{align}
Thus we see that the amplitude in the electric line basis consists of factors of $(-1)^{\text{framing}(w_i)}$, which ensure that $i=1,2$ are both fermions, as well as a factor of $(-1)^{\text{link} (v_1, \partial v_2)}$, which ensures that the two fermions braid non-trivially with each other (which means also that their fusion product is a fermion).  This is precisely the braiding amplitude of the $3$-fermion theory, or the ground state amplitude of the corresponding Walker-Wang model.

Now let us use the same formalism to determine the electric basis amplitudes of $\Ufr |\psi_{\text{pr}}\rangle$.  This calculation involves only a single copy of the $\{1,f\}$ Walker-Wang Hilbert space, so there is no more index $i$.  The same arguments as above lead to the expression 

\begin{align}
\langle \{v_{e}\} |\Ufr |\psi_{\text{pr}}\rangle = C'' \sum_{w_{e}} (-1)^{\text{framing}(\partial w) + \text{int}(w, v)  + \text{framing}(v)}
\end{align}
Now, we can write $(-1)^{\text{framing}(\partial w)} = (-1)^{\text{int}(w, \partial w)}$, where because $w$ and $\partial w$ now live on the same lattice we define the intersection by first translating $w$ by $(\frac{1}{2},\frac{1}{2},\frac{1}{2})$.  Shifting the variable $w\rightarrow w+y$ now results in 

\begin{align*}
\text{int}(w, \partial w) \rightarrow \text{int}(w, \partial w) + \text{int}(w, D(\partial y)) + \text{framing}(\partial y)
\end{align*}
where $D$ acts on closed homologically trivial $1$-chains $z$ by `doubling' them.  That is, $D(z) = z + T(z)$, where $T$ is the translation in the $(1,1,1)$ direction.

Now let us assume that the dimensions of the $3$-torus are all odd and relatively prime.  Then we claim that $D$ is invertible on closed homologically trivial $1$-chains.  To see this, note that the kernel of $D$ consists of $z$ that are translationally invariant in the $(1,1,1)$ direction.  Because the dimensions of the torus are relatively prime, this means that for each type of link $(x,y,z)$, they must all be occupied or all unoccupied.  Let us focus first on the $x$ links.  If they are all occupied then all the homologically non-trivial lines in the $x$ direction are occupied.  But the number of these lines is the product of the $y$ and $z$ dimensions of the torus, which is odd.  This is a homologically non-trivial configuration, and hence not allowed.  Similarly the links in the $y$ and $z$ directions cannot be occupied, so the kernel consists of just the empty configuration, i.e. $D$ is invertible.

This means that we can choose $y$ such that $\partial y = D^{-1}(v)$, which cancels the linear term and results in 
\begin{align}
\langle \{v_{e}\} |\Ufr |\psi_{\text{pr}}\rangle = C''' (-1)^{\text{framing}(D^{-1}(v)) +  \text{framing}(v)}
\end{align}
Unfortunately there does not seem to be a simple topological interpretation for $\text{framing}(D^{-1}(v))$.  Indeed, even for simple short closed $1$-cocycles, the application of $D^{-1}$ yields complicated nearly space-filling curves.

\section{Outlook: Two Semion QCA and Higher Dmensions}
\label{semionQCA}
\subsection{Two Semion QCA}
We have given a particularly simple form of the $3$ fermion QCA as conjugation by $\Ufr$, up to a circuit, or alternatively as conjugation by $U_{\text{link}}$.
The calculation of \cref{app:trivialize} suggests an interesting way to think about {\emph{why}} these QCAs are nontrivial.  Namely: ``they would be trivial if we had access to the gauge fields, but we do not".  For example, suppose we had two bosonic $\mathbb{Z}_2$ gauge theories on interpenetrating lattices.  Each theory has qubit degrees of freedom on links of some lattice, with the Pauli $Z$ operator on a link called a ``gauge field", and the product of gauge fields around a link being called a ``gauge flux".  Then, a unitary given by $(-1)$ raised to the power of the linking of the two gauge flux configurations is a circuit, as this is equivalent to $(-1)$ raised to the power of the intersection of type $1$ gauge fields with type $2$ gauge fluxes, or vice versa.  This gives us a circuit representation, indeed as a product of controlled-$Z$ gates.

However, the individual gates in this circuit representation do not respect gauge invariance, i.e., they do not commute with vertex operators which are products of Pauli $X$ on edges incident to a vertex.  
Indeed, any way to write $(-1)^{\text{linking}}$  as a circuit with gauge invariant gates should immediately give a way to write $U_{\text{link}}$ for the bosonized fermionic gauge theory as a circuit also, so
if indeed
conjugation by $U_{\text{link}}$ is a nontrivial QCA, then conjugation by
$(-1)^{\text{linking}}$ for the bosonic gauge theory is nontrivial as a $1$-form symmetry protected QCA, i.e., it has no representation as a shallow depth circuit if we require that the gates respect gauge invariance.

In our case, with the Walker-Wang model, we do not impose any such gauge invariance requirement on the circuit, but the gauge fluxes are {\emph{not}} products of some set of ``gauge fields" which commute with each other; rather, bosonization means that the operators we have access to in the qubit theory are products of gauge fields times Majorana hopping operators, and these operators do not commute with each other.  So, this has a similar effect to imposing gauge invariance.

Given any QCA for a bosonic gauge theory protected by $1$-form symmetry, then it is natural to define analogous QCAs for the gauge fields arising from bosonization.  
For example, consider the unitary $U_{\text{semion}}$ (on a single copy of the bosonized theory, i.e. a single copy of the $\{1,f\}$ Walker-Wang model) which multiplies any configuration of $\Z_2$ gauge flux lines $L$ by the braid amplitude for a set of semionic world lines following $L$.  This is a local QCA by 
roughly the same argument as the ``direct argument" in \cref{localitypreservingsubsec}: under a local change in flux, the change in the braid amplitude can be
computed locally.

%Acting on the product state, this should produce a state where the flux lines have semionic statistics while the fermions braid nontrivially with them, i.e., it should produce the $\{1,s,f,fs\}$ Walker-Wang model.  Since $fs$ is also a semion with the same topological spin as $s$, this is equivalent to two copies of the ${1,s}$ Walker-Wang model.

The operator $U_{\text{semion}}$ acts on a bosonic Hilbert space which we interpret as the Hilbert space of the $\{1,f\}$ Walker-Wang model.  Let us examine the action of $U_{\text{semion}}$ on the trivial product state where all the electric $f$ lines are confined.  Given the complicated nature of the same state under the action of $\Ufr$, as discussed in \cref{Uact}, we anticipate that the answer will be similarly un-enlightening here.  However, just as in the case of $\Ufr$, the situation simplifies when we introduce two copies of the $\{1,f\}$ Walker-Wang model.  Indeed, we will see that in this situation $(1\otimes U_{\text{semion}})U_{\text{link}}$ produces the $\{1,{\bar{s}}\} \times \{1,{\bar{s}}\}$ Walker-Wang ground state when acting on a trivial product state.  Here $s$ has topological spin $i$ and ${\bar{s}}$ has topological spin $-i$.  Since $U_{\text{link}}$ is, up to a circuit, just $\Ufr$, this means that, again up to a circuit, $(1\otimes U_{\text{semion}})$ produces the $\{1,s\} \times \{1,s\}$ Walker-Wang ground state.

To see that $(1\otimes U_{\text{semion}})U_{\text{link}}$ produces the $\{1,{\bar{s}}\} \times \{1,{\bar{s}}\}$ Walker-Wang ground state when acting on a trivial product state $|\psi_{0,0}\rangle$, we first note that $U_{\text{link}}|\psi_{0,0}\rangle$ is a superposition of tensor products of all states where the electric flux configuration in copy $1$ is identified with the magnetic flux configuration in copy $2$, weighted by $(-1)^{\text{framing}}$.  Acting with $U_{\text{semion}}$ on the second copy turns those magnetic fluxes - and hence the electric fluxes of type $1$ - into semions.  Expressing everything in the electric flux basis, we see that the electric fluxes of type $1$ are semions, those of type $2$ are fermions, and they have mutual braiding.  This means that the bound state of the type $1$ and type $2$ electric fluxes are again semions of the same topological spin, so we have the desired $\{1,{\bar{s}}\} \times \{1,{\bar{s}}\}$ Walker-Wang ground state.

We now make the above sketch more formal, as follows.  We start with eq. \ref{eq:Ustate}:

\begin{align} \label{eq:Ustate1}
U_{\text{link}} |\psi_{\text{prod}}\rangle = C' \sum_{\{M_i\}} (-1)^{\text{int}(M_1,\partial M_2)} Z_{M_1} Z_{M_2} |\psi_{0,0}\rangle
\end{align}
We then note that

\begin{align*}
Z_{M_1} Z_{M_2} |\psi_{0,0}\rangle = D\sum_{{\text{closed}} E_1} (-1)^{\text{int}(M_1,E_1) + \text{framing}(E_1)} |E_1\rangle Z_{M_2}|\psi_0\rangle
\end{align*}
where $D$ is some constant and $E_1$ a configuration of closed electric flux loops in copy $1$.  This is just expressing the magnetic state $Z_{M_1}|\psi_0\rangle$ in the first copy in the electric basis.  Plugging this in to eq. \ref{eq:Ustate1}, we obtain

\begin{align*}
U_{\text{link}} |\psi_{\text{prod}}\rangle = C' \sum_{M_1,M_2,E_1} (-1)^{{\text{int}}(M_1,\partial M_2) + {\text{int}}(M_1,E_1) + {\text{framing}}(E_1)} |E_1\rangle Z_{M_2}|\psi_0\rangle
\end{align*}
where $C'$ is another constant.  Performing the sum on $M_1$ yields a delta function that sets $E_1 = \partial M_2$, in particular ensuring that it is homologically trivial:

\begin{align*}
U_{\text{link}} |\psi_{\text{prod}}\rangle = C'' \sum_{M_2} (-1)^{{\text{framing}}(\partial M_2)} |E_1 = \partial M_2\rangle Z_{M_2} |\psi_0\rangle
\end{align*}
Thus we have 

\begin{align*}
(1\otimes U_{\text{semion}})U_{\text{link}} |\psi_{\text{prod}}\rangle = C'' \sum_{M_2} (-i)^{{\text{framing}}(\partial M_2)} |E_1 = \partial M_2\rangle Z_{M_2} |\psi_0\rangle
\end{align*}
Writing $Z_{M_2} |\psi_0\rangle$ in the electric basis we then obtain

\begin{align*}
(1\otimes U_{\text{semion}})U_{\text{link}} |\psi_{\text{prod}}\rangle &= C'' \sum_{M_2, E_2} (-i)^{{\text{framing}}(\partial M_2)} (-1)^{\text{int}(M_2,E_2) + \text{framing}(E_2)} |E_1 = \partial M_2\rangle |E_2\rangle
\end{align*}
Finally, we can re-write the sum over $M_2$ as the sum over $E_2 = \partial M_2$ at the expense of another overall constant and the sum over $E_2$ becoming a sum over homologically trivial $E_2$ (homologically non-trivial $E_2$ have $0$ amplitude because of interference between the different homology classes of $M_2$).  Thus

\begin{align*}
(1\otimes U_{\text{semion}})U_{\text{link}} |\psi_{\text{prod}}\rangle &= C''' \sum_{E_1, E_2} (-i)^{{\text{framing}}(E_1)} (-1)^{\text{link}(E_1,E_2) + \text{framing}(E_2)} |E_1\rangle |E_2\rangle
\end{align*}
where the sum is over closed, homologically trivial $E_1$ and $E_2$.  This is just the ground state wave function of the $\{1,{\bar{s}}\} \times \{1,{\bar{s}}\}$ Walker-Wang model.  $E_1$ is identified with the first semion, and $E_1 E_2$ with the second semion; $E_2$ is a fermion.

It is natural to conjecture that conjugation by $U_{\text{semion}}$ is (up to a circuit) the {\emph square} of the QCA of \cite{Shirley} which creates a single copy of the ${1,s}$ Walker-Wang model acting on a trivial product state.  After all, the square of a QCA is equal to (up to a circuit) two copies of that QCA, and two copies of the QCA of \cite{Shirley} produces the same Walker-Wang model as $U_{\text{semion}}$  does.  However, this is a conjecture as we do not know that the action on other states is the same.
At the same time, one may verify that the three-fermion QCA is the square of conjugation by $U_{\text{semion}}$, up to a circuit, by computing the square of
the semion braid amplitude.

\subsection{Higher Dimensions}
The form $U_{\text{link}}$ suggests a natural generalization to higher dimensions.  Consider any odd dimension $d=2k+1$.  
Consider $k+1$ different hypercubic lattices of qubits, each lattice slightly displaced in some generic direction, and apply the bosonization duality of
\cite{Kapustin3d} to each lattice.  Then, we have $k+1$ different magnetic fluxes, on copies labelled $1,2,\ldots,k+1$.
Each magnetic flux is a $2$-cocyle.  Dually, magnetic fluxes are $(d-2)$-cycles.
The intersection of magnetic fluxes $2,3,\ldots,k+1$ is a $1$-cycle that we call $C$.
Then, consider a higher dimensional unitary $U_{\text{higher}}$ that applies a phase equal to $-1$ to the linking number of the type $1$ flux with the chain $C$.

Equivalently, the type $1$ magnetic flux in this dual picture is the boundary of some closed $(d-1)$-cycle $F$.   Then, the phase is equal to $-1$ to the intersection number of $F$ with $C$, or equivalently the intersection of $F$ with fluxes $2,3,\ldots,k+1$.  Since $C$ is closed, this intersection number is the same for all homologically equivalent $F$.

We can immediately establish that $U_{\text{higher}}$ is a QCA by the same  ``direct argument" in \cref{localitypreservingsubsec}: under a local change in flux, the change in phase can be computed locally.

This form of $U_{\text{higher}}$ is reminiscent of higher dimensional Chern-Simons theory.  Indeed, perhaps this is no surprise: both nontrivial QCAs and Chern-Simons theory are related in that they both involve quantities that cannot be computed locally but whose variation can be computed locally.

We conjecture that $U_{\text{higher}}$ is nontrivial.  It would be interesting to understand if there is any relation between $U_{\text{higher}}$ and the nontrivial Clifford QCAs of \cite{haah2022topological} for qubits in odd dimensions $\geq 3$.   The expression we have given for $U_{\text{higher}}$ is not Clifford for $d\geq 5$, and these QCAs might not be circuit equivalent.  If so, $U_{\text{higher}}$ would represent a new class of QCA in odd dimension $5$ and higher.
It is possible that the Clifford QCAs of \cite{haah2022topological} are related to higher form Chern-Simons theory.

\subsection{Beyond Cohomology Phases}
Finally, it is interesting to speculate whether the simplified form $U_{\text{link}}$  can help simplify the construction of \cite{fidkowski2020exactly}.  There, a model for a $4+1$ dimensional beyond cohomology phase was constructed, whose boundary action was the three-fermion QCA.  However, the construction was quite complicated as it required decorating three-dimensional domain walls (of four-dimensional Ising degrees of freedom) with a three-fermion Walker-Wang model, and hence required defining the three-fermion Walker-Wang model and QCA on arbitrary closed three-dimensional geometries.  We conjecture that one can define a simpler theory whose boundary action is in the same phase as follows.  Take two hypercubic lattices of qubits in four dimensions, slightly displaced from each other, and apply the bosonization duality of
\cite{Kapustin3d} to each lattice, giving two gauge fields, labelled $1,2$.  We can regard the fluxes as $2$-cycles.  Take one more hypercubic lattice of qubits, again displaced from the others, and call the qubits on this lattice the ``Ising" degrees of freedom.  The domain walls of the Ising degrees of freedom give a $3$-cycle.  The intersection of this $3$-cycle with the type $2$ flux is a $1$-cycle we call $C$.  Define a unitary $V$ which is equal to $-1$ to the linking number (in four dimensions) of type $1$ flux with this cycle $C$.  Then, $V^2=1$, and we regard $V$ as the disentangler for a symmetry protected phase which has an Ising symmetry.  This symmetry flips all Ising degrees of freedom, and hence commutes with $V$ on any closed manifold.

{\bf{Acknowledgements:}} LF would like to acknowledge support from the National Science Foundation under grant DMR-2300172.

\appendix

\section{Trivializing the QCA By Tensoring With Fermions}
\label{app:trivialize}
We have shown that the 3F QCA is a circuit if we tensor with the identity QCA acting on fundamental fermions.
In this appendix, we give a particularly simple form for this circuit, writing it as a product of gates which commute with each other.

Before giving this simple form, let's first give a brief abstract argument that it trivializes with fundamental fermions.  The 3F QCA $\alpha_{3F}$, corresponding to the unitary $U_\text{link}$ squares to the identity.  So, $\alpha_{3F} \otimes \alpha_{3F}$ is a circuit acting on two copies of the system.  Suppose on the second copy of the system, there is no gauge flux.  Then, the QCA $\alpha_{3F}$ act trivially on that copy, and so the action of $\alpha_{3F} \otimes \alpha_{3F}$ is the same as $\alpha_{3F}$ tensored with the identity QCA.  However, tensoring with a copy of the system constrained to have no gauge flux is the same as tensoring with fundamental fermions.

\subsection{Trivializing the QCA}
We trivialize the unitary $U_\text{link}$ in the presence of fundamental fermions.
Recall that under the bosonization duality of \cite{Kapustin3d}, the qubit system is dual to one where each $3$-cell $c$ has two Majorana operators, $\gamma_c$ and $\gamma_c'$.  Gauge fields live on $2$-cells, while there is gauge flux on $1$-cells.  The operator $i\gamma'_c \gamma_c$ is gauge invariant.
Other gauge invariant operators include the operator $i\gamma_c \gamma_d$, for two different $3$-cells $c,d$ connected by a face, multiplied by the gauge field on a face, and so these operators have an image as some local bosonic operator.  Let us denote that bosonic operator by $U_{c,d}$ for any two $3$-cells $c,d$ which are incident on some $2$-cell.

The gauge flux on a $1$-cell is equal to, up to a sign, $U_{c,d} U_{d,e} U_{e,f} U_{f,c}$, where $c,d,e,f$ are the four different $3$-cells incident to that $1$-cell, taken in order going around the $1$-cell in an arbitrary direction, so that $c,d$ are incident on a $2$-cell as needed to define $U_{c,d}$. and similarly for the other pairs.

The set of operators $\{U_{c,d}\}$ do not commute with each other.  Now, we tensor in ancilla fundamental fermionic degrees of freedom, by tensoring in operators $\eta,\eta'$ on the $3$-cells of one of the two cubic lattices which we will arbitrarily pick to be lattice $1$.

We consider the unitary $U_\text{link}$ tensored with the identity on the ancilla fermions, and will show that it can be represented by a circuit.
Remark: in fact we will only need the degrees of freedom $\eta$, and not those $\eta'$ to do this.

Define the operator $T_{c,d}$ by
$$T_{c,d}=i U_{c,d} \eta_c \eta_d.$$
Now, the set of operators $\{T_{c,d}\}$ enjoy the following properties: they are mutually commuting, they have eigenvalues $\pm 1$, and for any $1$-cell, the flux on that $1$-cell is 
equal to, up to a sign, $T_{c,d} T_{d,e} T_{e,f} T_{f,c}$, where $c,d,e,f$ are four different $3$-cells incident to that $1$-cell as before.

The fact that they are mutually commuting may be explicitly checked, but the intuitive explanation is that two operators $U_{c,d}$ and $U_{g,h}$ have been constructed to reproduce the Majorana anti-commutation relations of operator $\gamma_c \gamma_d$ with $\gamma_g \gamma_h$ so that they anti-commute if the set $\{c,d\} \cap \{e,f\}$ has one element.  The fact that the eigenvalues are $\pm 1$ follows since each $U_{c,d}$ has eigenvalues $\pm 1$, as does $i\eta_c \eta_d$.  The fact that
the gauge flux is equal to $T_{c,d} T_{d,e} T_{e,f} T_{f,c}$ up to sign follows because the product $(\eta_c \eta_d) (\eta_d \eta_e) (\eta_e \eta_f) (\eta_f \eta_c)$ equal $+1$.

Since the $T_{c,d}$ mutually commute, we may work in a simultaneous eigenbasis of these operators.  Regard these operators as defining some ``type $1$ gauge field", as the desired type $1$ gauge flux is computed from their products.  
Then the mod $2$ linking number of type $1$ flux with type $2$ flux is equal to the mod $2$ number of the type $1$ gauge field with the type $2$ gauge flux.
Precisely, the linking number is equal, mod $2$, to the number of type $2$ $1$-cells which have gauge flux $-1$ and which intersect a type $1$ $2$-cells which has gauge field $-1$.

Then, $-1$ to the linking number is equal to
a product of local gates, one local gate for each $2$-cell in the type $1$ lattice.  These gates are diagonal in an eigenbasis of the $T_{c,d}$ and the type-$2$ gauge flux operators. Each gate gives a $-1$ phase if the gauge field on that cell equals $-1$ and if the gauge flux on the cell in the type $2$ lattice intersecting it also equals $-1$; otherwise the gate gives a $+1$ phase.

We can write this in a more symmetric form if we also introduce Majorana operators on the $3$-cells of the type-$2$ lattice.  Then, let the ``type-$1$ gauge field" $T^{(1)}_{c,d}$ be the operators $T_{c,d}$ above and let  the ``type-$2$ gauge field" $T^{(2)}_{c,d}$ be analogous operators for the type-$2$ lattice.  Then, since the gauge flux for the type-$2$ lattice can be written as a product of operators $T^{(2)}_{c,d}$, the circuit assumes a more symmetric form:
it is a product over all pairs $(c^{(1)},d^{(1)})$ and $(c^{(2)},d^{(2)})$, where $c^{(1)},d^{(1)}$ are in the type-$1$ lattice and $c^{(2)},d^{(2)}$, such that the boundary of the face $(c^{(1)},d^{(1)})$  intersects the face $(c^{(2)},d^{(2)})$, of a ``controlled-$Z$" gate.  This controlled-$Z$ gate is diagonal in the eigenbasis of the gauge fields and gives a $-1$ phase if both gauge fields equal $-1$, and otherwise gives a $+1$ phase.  Note that while our definition is superficially not symmetric in the lattices $1,2$, as we considered the intersection of the boundary of a type-$1$ face with a type-$2$ face, this is actually equivalent to considering the intersection of the boundary of a type-$2$ face with a type-$1$ face, so it is a completely symmetric definition.

\subsection{Trivial Circuit?}
Next we ask: is this circuit a \emph{trivial circuit}?  There are several possible definitions of a trivial circuit, and of a \emph{trivial symmetry}, and we now discuss these.

Let us say a QCA or circuit is a symmetry if it gives a (possibly projective) representation of some symmetry group\footnote{In the case of a QCA, there is no meaning to whether or not the representation is projective, as multiplying a state by a phase corresponds to the identity QCA.}, i.e., if there is some mapping from group elements $g$ to unitary circuits $U(g)$ or QCA $\alpha(g)$ giving a group homomorphism up to phase.
In this case, the QCA squares to the identity and the symmetry is $\mathbb{Z}_2$.

The definition of a trivial circuit that we use in this subsection is that it can be conjugated by some unitary circuit so that the result can be decomposed as a product of unitary gates, which have disjoint support, and
the definition of a trivial symmetry that we use is that there is some single unitary circuit (independent of the group element $g$) so that each $U(g)$ can be conjugated by that unitary circuit
so that the result can be decomposed as a product of local unitary gates, each of which gives a (possibly projective) representation of that symmetry and which have disjoint support, i.e., each gate may act on more than one site, but the support of the gates must be disjoint from each other so that this can be expressed as a quantum circuit with depth $1$.

We will show that our circuit is not a trivial circuit under this definition.  More strongly, we will show that the image of our circuit under an arbitrary QCA
cannot be a product of local unitary gates with disjoint support.
Interestingly, if one regards this circuit as the boundary symmetry action of some SPT with a $\mathbb{Z}_2$ symmetry, and applies 
the classification method of \cite{else2014classifying}, no obstruction is detected.

Note that a possible weaker definition of a trivial circuit is that it 
can be conjugated by some unitary circuit so that the result can be decomposed as a
a product of local unitary gates which commute with each other.  A weaker definition of a trivial symmetry is that there is some single unitary circuit (independent of the element $g$ of the symmetry group) so that each $U(g)$ 
can be conjugated by some unitary circuit so that the result can be decomposed as a product of local unitary gates which commute with each other and such that each gives a (possibly projective) representation of that symmetry.
Under the weaker definition, our circuit is a trivial symmetry; indeed it already has that decomposition without needing to conjugate by anything.

Note also that our definition of a trivial circuit is slightly weaker than an alternative definition where one requires that the unitary gates each act on a single site.

We give a simpler result first: we show that the conjugation cannot be done in a translationally invariant manner on a three-torus with some finite unit cell (i.e., the translation symmetry group may be larger than translation by a single lattice site).  Indeed, suppose some such conjugation could be done.   Consider the normalized trace of the circuit, defined to be the trace of the given unitary divided by the trace of the identity.  Then, if it could be conjugated to a product of unitaries with disjoint support, the normalized trace would be the product of traces of these unitaries and so would have an exactly exponential dependence on volume if the linear size is a multiple of the unit cell for translation symmetry.
However, we show next that the normalized trace
equals $2^{b_1-b_0}$ times an exponential function of volume, where $b_1=3$ and $b_0=1$ are Betti numbers of the three-torus.  This gives a contradiction.
 
 We compute the normalized trace by first averaging the sign of the unitary over type $1$ gauge field for fixed type $2$ gauge field, and then averaging over type $1$ gauge field.  If the type $2$ gauge flux is nonzero anywhere, then the average over type $1$ flux vanishes, while if the type $2$ gauge flux is zero everywhere, then the sign of the unitary is $+1$ independent of the type $1$ gauge flux.  So, the normalized trace is the probability that the type $2$ gauge flux vanishes everywhere.  This is equal to the number of type $2$ gauge field configurations with no gauge flux, divided by the total number of type $2$ gauge field configurations.  The number of gauge field configurations with no gauge flux is equal to the number of homology classes (which equals $2^{b_1}$) times the number of inequivalent gauge transformations (which equals $2^{n_V-b_0}$).  So, indeed, the normalized trace is $2^{b_1-b_0}$ times an exponential function.
 
 Remark: this argument that there is no translation invariant trivialization does not use the presence of fermions in any way.  It would work if we had, for example, qubits on each link of the type $1$ and type $2$ lattice, with the value of the qubit giving the gauge field, and compute the linking number of the corresponding gauge flux.
 
 Remark: also note that a similar argument works if we consider a circuit in one dimension given by a product of controlled-$Z$ gates on all nearest neighbor pairs, either on a ring or on an interval.  Then, averaging over qubits on the even sublattice for fixed $Z$ configuration on the odd sublattice, the average vanishes unless all odd sublattice qubits are in the same $Z$ state, and there are $2$ such configurations.
 
 Now we show that the it cannot be done in a translationally non-invariant manner.  We show this just in the case of the one-dimensional circuit of the above paragraph; the three-dimensional circuit can be handled with essentially the same argument after dimensionally reducing to one dimension by ignoring two of the directions.
 
Consider the system on a ring.  Let $U$ be the circuit which is the product of controlled-$Z$ gates on nearest neighbors.  Let $V$ be the 
hypothetical circuit such that $VUV^\dagger$ a product of unitaries with disjoint support (the argument where we map by a QCA is similar; indeed, in one-dimension every QCA is a circuit composed with a shift so showing it in the case of a circuit suffices).
Divide the ring into four disjoint intervals, called $A,B,C,D$ in order around the ring, with $A$ and $D$ neighbors and with the size of each interval long compared to the range of $V$ and to the range of the gates in $V U V^\dagger$.

Consider ${\rm Tr}_{B,D}(U)$, where ${\rm Tr}_{\ldots}(\cdot)$ denotes a partial trace over some region.  This partial trace does \emph{not} factorize into a product of operators supported on $A$ and $C$.  To see this, note that if we trace over qubits on the even sublattice in $B,D$, this requires qubits in the odd sublattce to agree in the $Z$ basis, and after tracing over qubits in the odd sublattice in $B,D$ it forces a qubit near the boundary $A,B$ to agree in the $Z$ basis with a qubit near the boundary $B,C$.

Now we show that if such a $V$ exists, then the partial trace ${\rm Tr}_{B,D}(U)$ would factorize, giving a contradiction.  Of course, the partial trace ${\rm Tr}_{B,D}(V U V^\dagger)$ does factorize by assumption that $VUV^\dagger$ is a product of unitaries with disjoint support, but this is not what we want to show.
To show what we want,
it is useful to define the notion of a trace over an algebra rather than a site: given any simple algebra ${\cal F}$, which is a subalgebra of the algebra of all operators on this system, the trace of an operator $O$ over that subalgebra can be defined by giving the full Hilbert space a tensor product structure
${\cal H}_1 \otimes {\cal H}_2$ such that ${\cal F}$ is the algebra of operators on ${\cal H}_1$, and then
tracing over ${\cal H}_1$.

Now, consider some subintervals $B'\subset B$ and $D'\subset D$ such that $B'$ is large compared to the size of the disjoint gates in $VUV^\dagger$ and such that the distance from $B'$ to the boundary of $B$ is large compared to the range of $V$ and similarly for $D',D$.  Let $A',C'$ be intervals such that $A\subset A'$ and $C\subset C'$ with $A',B',C',D'$ giving some disjoint decomposition of the ring.
We have that
${\rm Tr}_{B',D'}(VUV^\dagger)$ factorizes into a product of operators on $A',C'$.
Now, the trace over $B'$ is the same as the trace over the algebra of operators on $B'$, which we call ${\cal B}'$.
By assumption on the range of $V$, the algebra $V^\dagger {\cal B}' V$ is a simple subalgebra of the algebra ${\cal B}$ of operators on $B$.  
Further, since $V$ is local, the
commutant of $V^\dagger {\cal  B}' V$ in ${\cal B}$
decomposes as a tensor product of two simple subalgebras, one supported near the boundary between $A$ and $B$ and the other supported near the boundary between $B$ and $C$.   Call these subalgebras ${\cal B}_{L}$ and ${\cal B}_R$, where the subscripts are for left, right.
So, ${\cal B}$ factorizes as a product of ${\cal B}_L$ and $V^\dagger {\cal B}' V$ and ${\cal B}_R$.
So, the trace of $U$ over $B$ is equal to the trace of $U$ over $V^\dagger {\cal B}' V$ and ${\cal B}_L$ and ${\cal B}_R$.
Similarly, the trace of $U$ over $D$ is equal to the trace of $U$ over $V^\dagger {\cal D}' V$ and ${\cal D}_L$ and ${\cal D}_R$, where we define ${\cal D}_L$ and ${\cal D}_R$ analogously to ${\cal B}_L$ and ${\cal B}_R$.
We may take the partial traces in any order, and we choose to trace over
$V^\dagger {\cal B}' V$ and
$V^\dagger {\cal D}' V$ first.
However, by the assumption that $VUV^\dagger$ is a product of disjointly supported gates, the partial trace of $U$ over $V^\dagger {\cal B}' V$ and
$V^\dagger {\cal D}' V$ factorizes into a product of two operators, one supported on the algebra generated by ${\cal A}$ and ${\cal B}_L$ and by ${\cal D}_R$, and
one supported on the algebra generated by ${\cal C}$ and ${\cal B}_R$ and by ${\cal D}_L$.
Then, tracing over ${\cal B}_L$, ${\cal B}_R$, ${\cal D}_L$, and ${\cal D}_R$, it follows that
the partial trace of $U$ over $B$ and $D$ factorizes, giving the desired contradiction.

\section{Details of the construction of the Chern insulator pump, and computation of its winding number} \label{app:Chern}

\subsection*{Single Chern insulator}

We first discuss a single Chern insulator in a $2$ dimensional 2 band model with only the flavor index $\alpha$.  We let $\vk=(k_x,k_y)$ be the $2d$ reciprocal wave-vector.

\begin{align*}
H_{\text{C.I.}} = c^X(\vk) X^\fl + c^Y(\vk) Y^\fl + c^Z(\vk) Z^\fl
\end{align*}
be a Hamiltonian for a Chern insulator with Chern number $\pm 1$.  That is, the coefficients $c^X(\vk), c^Y(\vk), c^Z(\vk)$ are chosen such that $(c^X)^2+ (c^Y)^2+(c^Z)^2=1$ and
the function $\vk \rightarrow (c^X,c^Y,c^Z)$, viewed as a map from $T^2$ to $S^2$, has winding number 1.  We will find it convenient to work with a specific form of this map, which we construct as follows.  Take $k_1 < k_2 \ll 1$.  Let $\theta(\vk)$ interpolate smoothly between $0$ for $|\vk|>k_2$ and $\pi$ for $|\vk| \leq k_1$, and define

\begin{align*}
(c^X,c^Y,c^Z) = \begin{cases}
(0,0,-1) & \text{for } |\vk| < k_1 \\
(0,0,1) & \text{for } |\vk| > k_2 \\
(\frac{k_x}{|\vk|} \sin(\theta(\vk)) , \frac{k_y} {|\vk|}\sin(\theta(\vk)), \cos(\theta(\vk))) & \text{for } k_1 \leq |\vk| \leq k_2
\end{cases}
\end{align*}
Now let $\phi(\vk) = \tan^{-1} \left(k_y / k_x\right)$, and, for $|\vk|>k_1$, define the operator

\begin{align*}
V_+(\vk) = \exp\left(i \phi(\vk) Z^\fl/2 \right) \exp\left(i \theta(\vk) Y^\fl/2\right)
\end{align*}
Note that the states $V_+(\vk)|\alpha=0,1\rangle$ are eigenstates of $H_{\text{C.I.}}(\vk)$:

\begin{align} \label{eq:conj}
H_{\text{C.I.}}(\vk) V_+(\vk)|\alpha\rangle &= (-1)^\alpha V_+(\vk)|\alpha\rangle
\end{align}
Similarly, for $|\vk|>k_1$, we define the operator
\begin{align*}
V_-(\vk) = \exp\left(-i \phi(\vk) Z^\fl/2 \right) \exp\left(i \theta(\vk) Y^\fl/2 \right)
\end{align*}
which has the same property for the Hamiltonian $c^X(\vk) X^\fl - c^Y(\vk) Y^\fl + c^Z(\vk) Z^\fl$, which has Chern number $-1$.

\subsection*{Nucleating and annihilating pairs of Chern insulators}

We now construct the operator $\Un(\veck)$ which nucleates pairs of Chern number $\pm 1$ insulators on layers $(2n,2n+1)$ (i.e. $(l=0,l=1)$).  We define
%Let us now nucleate pairs of Chern number $\pm 1$ insulators on layers $(2n,2n+1)$ (i.e. $(l=0,l=1)$).  This will be described by an operator ${\Un}(\veck)$, with $\veck=(k_x,k_y,k_z)$ the $3$ dimensional reciprocal lattice vector.  We will argue that the ${\Un}(\veck)$ we construct can be smoothly connected to the identity, as a function of $\veck$, essentially because the bundle that describes a pair of $+1$,$-1$ Chern number bands is trivial.  Hence ${\Un}(\veck)$ can be generated by a finite time evolution of a quasi-local Hermitian operator.  Now, since we nucleate the exact same pairs of Chern insulators on each pair of layers $(2n,2n+1)$, without coupling across the (doubled) unit cells in the $z$ direction, ${\Un}(\veck)$ will be independent of $k_z$.  We thus have to define the $4$ by $4$ unitary matrix ${\Un}(\veck)$ as function of $\vk$.  In fact, we will ensure that ${\Un}(\veck) \in SU(4)$.

\begin{align} \label{eq:Udef}
{\Un}(\veck) = \begin{cases}
V_+(\vk)(1+Z^\lay)/2 + V_-(\vk)(1-Z^\lay)/2 & \text{for } |\vk| \geq k_1\\
\left(|\vk|/k_1\exp\left(i\phi(\vk)Z^\lay Z^\fl\right) + i \sqrt{1-|\vk|^2/k_1^2} \,X^\lay\right)i Y^\fl & \text{for } |\vk| < k_1
\end{cases}
\end{align}
Note that for $|\vk| = k_1$ we have $\theta(\vk)=\pi$, so ${\Un}(\veck)$ reduces to $\exp\left(i\phi(\vk)Z^\lay Z^\fl / 2\right) i Y^\fl$ on this circle.  This means that the map defined above is continuous.  It is not smooth, but can presumably be deformed into a smooth map.  We claim that the operator ${\Un}(\veck)$ satisfies ${\Un}^\dagger H_{\text{pair}} \Un = Z^\fl$, i.e. it nucleates a pair of Chern number $\pm 1$ Chern insulators.  For $|\vk| > k_1$ this follows from eqs. \ref{eq:conj} (and the corresponding equation for $V_-$) and \ref{eq:Udef}.  For $|\vk| < k_1$ we have both $H_{\text{C.I.}} = -Z^\fl$ and $H_{\text{pair}} = -Z^\fl$, and again using eq. \ref{eq:Udef} we see $Z^\fl = {\Un}(\veck)^\dagger H_{\text{pair}} {\Un}(\veck)$.

We now annihilate complementary pairs of Chern insulators.  The operator $\Ua(\veck)$ which does this can be constructed as the conjugation by a translation by $1$ in the $z$ direction of an operator that performs the annihilation on the same pairs of layers as the nucleation.  Viewed in terms of the doubled unit cell, this translation by $1$ is accomplished by swapping the layer index, and then translating one of the layers by $2$ (i.e. by a single doubled unit cell); this operator is $X^\lay \exp\left(ik_z(1-Z^\lay)/2\right)$.  On the other hand, the operator that performs the annihilation is just $X^\lay {\Un}(\veck)^{-1} X^\lay$.  We thus have

\begin{align*}
\Ua(\veck) &= \left(X^\lay \exp\left(ik_z(1-Z^\lay)/2\right)\right)^{-1}  \left(X^\lay \Un(\veck)^{-1} X^\lay\right) \left(X^\lay \exp\left(ik_z(1-Z^\lay)/2\right)\right)\\
&= \exp\left(-ik_z(1-Z^\lay)/2\right) \Un(\veck)^{-1} \exp\left(ik_z(1-Z^\lay)/2\right)
\end{align*}
Therefore

\begin{align*}
U(\veck) = \Ua(\veck) \Un(\veck) &= \exp\left(-ik_z(1-Z^\lay)/2\right) \Un(\veck)^{-1} \exp\left(ik_z(1-Z^\lay)/2\right) \Un(\veck) \\
&= \exp\left(ik_z Z^\lay/2\right) \Un(\veck)^{-1} \exp\left(-ik_z Z^\lay/2\right) \Un(\veck)
\end{align*}
The operator $U(\veck)$ can again be continuously connected to $\Un(\veck)$ by the same argument as before, so is a finite time evolution of a quasi-local Hamiltonian.  Although it preserves the ground state of $H_{\text{triv}}$, it is not the identity operator.  We now analyze its structure.

First, note that since for $|\vk| \geq k_1$ $\Un(\veck)$ is block diagonal in layer space, i.e. it commutes with $Z^\lay$, so that $U(\veck)=1$ for $|\vk| \geq k_1$.  For $|\vk| < k_1$ we have, for $r\equiv |\vk|/k_1$:

\begin{align*}
U(\veck) &= \exp\left(ik_z Z^\lay/2\right)Y^\fl \left(r \exp\left(i \phi Z^\lay Z^\fl / 2\right) + i\sqrt{1-r^2}\,X^\lay\right)^{-1} \exp \left(-ik_z Z^\lay/2\right) \left(r \exp\left(i \phi Z^\lay Z^\fl/2 \right) + i\sqrt{1-r^2}\,X^\lay\right) Y^\fl \\
&=\exp\left(ik_z Z^\lay/2\right)Y^\fl \left(r \exp\left(-i \phi Z^\lay Z^\fl/2 \right) - i\sqrt{1-r^2}\,X^\lay\right)\exp \left(-ik_z Z^\lay/2\right)\left(r \exp\left(i \phi Z^\lay Z^\fl /2 \right) + i\sqrt{1-r^2}\,X^\lay\right) Y^\fl \\
&=\exp\left(ik_z Z^\lay/2\right) \left(r \exp\left(i \phi Z^\lay Z^\fl /2 \right) - i\sqrt{1-r^2}\,X^\lay\right)\exp \left(-ik_z Z^\lay/2\right)\left(r \exp\left(-i \phi Z^\lay Z^\fl/2 \right) + i\sqrt{1-r^2}\,X^\lay\right) \\
&=\left(r \exp\left(i \phi Z^\lay Z^\fl /2 \right) - i\sqrt{1-r^2}\,\left( \cos k_z X^\lay + \sin k_z Y^\lay \right)\right)\left(r \exp\left(-i \phi Z^\lay Z^\fl /2 \right) + i\sqrt{1-r^2}\,X^\lay\right)
\end{align*}
Since this operator commutes with $Z^\fl$, we can focus on one of the $Z^\fl$ eigenvalues, say $Z^\fl=1$.  Then 
\begin{align*}
U(\veck) &= \left(r \exp\left(i \phi Z^\lay / 2 \right) - i\sqrt{1-r^2}\,\left( \cos k_z X^\lay + \sin k_z Y^\lay \right)\right)\left(r \exp\left(-i \phi Z^\lay/2 \right) + i\sqrt{1-r^2}\,X^\lay\right) \\
&= \left(r^2 + (\cos k_z)(1-r^2)\right) \cdot 1 + a_X X^\lay + a_Y Y^\lay + a_Z Z^\lay
\end{align*}
where $a_X,a_Y, a_Z$ are coefficients whose precise form will not be necessary.  From the above expression, we see that $U(\veck)=-1$ if and only if $r=0$, i.e. $\vk = 0$ and $k_z=\pi$.  That is, $U(\veck)=-1$ only for $\veck = (0,0,\pi)$.  Hence the map from $T^3$ to $SU(2)$ defined by $U(\veck)$ for $Z^\fl=1$ has winding number, or degree, $1$.  Similarly, the map from $T^3$ to $SU(2)$ defined by $U(\veck)$ for $Z^\fl=-1$ has winding number $-1$.

According to the definition in the main text, we have
\begin{align*}
\Ua(\veck) &= \left(X^\lay \exp\left(ik_z(1-Z^\lay)/2\right)\right)^{-1}  \left(X^\lay \Un(\veck)^{-1} X^\lay\right) \left(X^\lay \exp\left(ik_z(1-Z^\lay)/2\right)\right)\\
&= \exp\left(-ik_z(1-Z^\lay)/2\right) \Un(\veck)^{-1} \exp\left(ik_z(1-Z^\lay)/2\right)
\end{align*}
Thus

\begin{align*}
U(\veck) = \Ua(\veck) \Un(\veck) &= \exp\left(-ik_z(1-Z^\lay)/2\right) \Un(\veck)^{-1} \exp\left(ik_z(1-Z^\lay)/2\right) \Un(\veck) \\
&= \exp\left(ik_z Z^\lay/2\right) \Un(\veck)^{-1} \exp\left(-ik_z Z^\lay/2\right) \Un(\veck)
\end{align*}
The operator $U(\veck)$ can again be continuously connected to $\Un(\veck)$ by the same argument as before, so is a finite time evolution of a quasi-local Hamiltonian.  Although it preserves the ground state of $H_{\text{triv}}$, it is not the identity operator.  We will presently analyze its structure.

First, note that since for $|\vk| \geq k_1$ $\Un(\veck)$ is block diagonal in layer space, i.e. it commutes with $Z^\lay$, so that $U(\veck)=1$ for $|\vk| \geq k_1$.  For $|\vk| < k_1$ we have, for $r\equiv |\vk|/k_1$:

\begin{align*}
U(\veck) &= \exp\left(ik_z Z^\lay/2\right)Y^\fl \left(r \exp\left(i \phi Z^\lay Z^\fl / 2\right) + i\sqrt{1-r^2}\,X^\lay\right)^{-1} \exp \left(-ik_z Z^\lay/2\right) \left(r \exp\left(i \phi Z^\lay Z^\fl/2 \right) + i\sqrt{1-r^2}\,X^\lay\right) Y^\fl \\
&=\exp\left(ik_z Z^\lay/2\right)Y^\fl \left(r \exp\left(-i \phi Z^\lay Z^\fl/2 \right) - i\sqrt{1-r^2}\,X^\lay\right)\exp \left(-ik_z Z^\lay/2\right)\left(r \exp\left(i \phi Z^\lay Z^\fl /2 \right) + i\sqrt{1-r^2}\,X^\lay\right) Y^\fl \\
&=\exp\left(ik_z Z^\lay/2\right) \left(r \exp\left(i \phi Z^\lay Z^\fl /2 \right) - i\sqrt{1-r^2}\,X^\lay\right)\exp \left(-ik_z Z^\lay/2\right)\left(r \exp\left(-i \phi Z^\lay Z^\fl/2 \right) + i\sqrt{1-r^2}\,X^\lay\right) \\
&=\left(r \exp\left(i \phi Z^\lay Z^\fl /2 \right) - i\sqrt{1-r^2}\,\left( \cos k_z X^\lay + \sin k_z Y^\lay \right)\right)\left(r \exp\left(-i \phi Z^\lay Z^\fl /2 \right) + i\sqrt{1-r^2}\,X^\lay\right)
\end{align*}
Since this operator commutes with $Z^\fl$, we can focus on one of the $Z^\fl$ eigenvalues, say $Z^\fl=1$.  Then 
\begin{align*}
U(\veck) &= \left(r \exp\left(i \phi Z^\lay / 2 \right) - i\sqrt{1-r^2}\,\left( \cos k_z X^\lay + \sin k_z Y^\lay \right)\right)\left(r \exp\left(-i \phi Z^\lay/2 \right) + i\sqrt{1-r^2}\,X^\lay\right) \\
&= \left(r^2 + (\cos k_z)(1-r^2)\right) \cdot 1 + a_X X^\lay + a_Y Y^\lay + a_Z Z^\lay
\end{align*}
where $a_X,a_Y, a_Z$ are coefficients whose precise form will not be necessary.  From the above expression, we see that $U(\veck)=-1$ if and only if $r=0$, i.e. $\vk = 0$ and $k_z=\pi$.  That is, $U(\veck)=-1$ only for $\veck = (0,0,\pi)$.  Hence the map from $T^3$ to $SU(2)$ defined by $U(\veck)$ for $Z^\fl=1$ has winding number, or degree, $1$.  Similarly, the map from $T^3$ to $SU(2)$ defined by $U(\veck)$ for $Z^\fl=-1$ has winding number $-1$.

\section{Pumping a $p+ip$ state gives a $CCZ$ gate: explicit free fermion computation} \label{app:CCZ}

%We start with a $2$ band particle conserving free fermion unitary $U$ whose winding number (or degree) over the BZ is $1$, such as $U_{\alpha=0}(\veck)$ constructed in eq. \ref{eq:Ufinal}.  Note that in eq. \ref{eq:Ufinal}, $U$ is a $4$ by $4$ 

%This is in some sense half of the unitary that pumps a Chern insulator, and we will see below that it pumps a $p+ip$ state.  For now, let us see that $U$ implements a $CCZ$ gate on a $T^3$.  We imagine an $L \times L \times L$ cubic lattice with $2$ orbitals per site.  Let us denote the creation and annihilation operators on site ${\vec{r}}$ by $a_{\vec{r},i}, a^\dagger_{\vec{r},i}$ where $i=1,2$ is the orbital index.

%Since $U$ has a quantized winding number, it cannot be written as a particle number conserving free fermion circuit.  But, we will see that it can be written as a fermion-parity conserving free fermion circuit.  We will have to show that changing all $3$ boundary conditions from periodic to anti-periodic gives a $-1$ in the action of the circuit on Fock space.

\subsubsection*{Boundary conditions}

Let us see how to implement a change in boundary conditions (from periodic to anti-periodic) in a simple toy example of a $1$ dimensional tight binding model on a ring, with $L$ sites and one orbital per site.  We set the lattice constant to $1$ for simplicity.  Let us work with the time ordered exponential of the integral of a local time dependent quasi-Hamiltonian $K$.  $K$ is some bilinear in the creation and annihilation operators $a_1,\ldots, a_L$ and $a_1^\dagger,\ldots, a_L^\dagger$.  When boundary conditions are periodic, this means that there is an extra $-1$ in front of the term that tunnels fermion parity between sites $L$ and $1$; in particular, $K$ is not translation invariant.  However, we can work with the modes

\begin{align*}
{\tilde{a}}_k = \sum e^{i k j} a_j
\end{align*}
with $k= \frac{\pi}{L}+ \frac{2\pi m}{L}$, $m=0,\ldots, L-1$.  Then by inverting this and plugging into $K$ we see that $K$ can be written as:

\begin{align*}
K = \sum_{k = \frac{\pi}{L}+ \frac{2\pi m}{L}} C_k {\tilde{a}}_k^\dagger a_k + \sum_{k = \frac{\pi}{L}+ \frac{2\pi m}{L}} D_k {\tilde{a}}_k {\tilde{a}}_{-k} + \text{h. c.}
\end{align*}
where $C_k$ and $D_k$ are some constants.  In other words, the effect of the $-1$ in the periodic boundary conditions is to shift the allowed quantized values of the reciprocal lattice wave vector by $\frac{\pi}{L}$.  For anti-periodic boundary conditions, on the other hand, we get the same expression except the sum is over $k = \frac{2\pi m}{L}$, $m=0,\ldots, L-1$.

In our three dimensional model of interest, the same shift in the allowed values of $\veck$ occurs.  In particular, it is only when all three boundary conditions are anti-periodic that we have values of $\veck$ for which $\veck = -\veck$ modulo the Brilliouin Zone, i.e. ones which are reflection symmetric.  We will see that only these points contribute to the desired sign.

\subsubsection*{General observation about free fermion unitary circuits}

Consider a general free fermion unitary evolution $V(\tau)=T\exp\left(i\int_0^\tau K(\tau') d\tau'\right)$, where $K(\tau')$ is quadratic in the creation and annihilation operators on $N$ sites.  It will now be useful to work with the Majorana representation of the operator algebra, where we trade the creation and annihilation operators for $2N$ Majorana modes $\gamma_1, \ldots, \gamma_{2N}$, defined by:

\begin{align*}
\gamma_{2j} &= a_j + a_j^\dagger \\
\gamma_{2j+1} &= i(a_j - a_j^\dagger)
\end{align*}
We will fix the overall additive constant in $K$ by demanding that $K=\frac{i}{4}\sum_{i,j} A_{ij} \gamma_i \gamma_j$, where $A_{ij}$ is anti-symmetric and $i$ and $j$ run from $1$ to $2N$.  Let $V(\tau) = T\exp\left(i\int_0^\tau K(\tau') d\tau'\right)$.  The $V(\tau)$ defines a path from $1$ to $V$ in the space of free fermion unitaries of determinant $1$ in the many-body Fock space.

The group of all free fermion unitaries of determinant $1$ on $N$ fermions is generated by all operators of the form $\exp\left(\frac{1}{4}\sum_{i,j} A_{ij} \gamma_i \gamma_j\right)$, where $A_{ij}$ is anti-symmetric and $i$ and $j$ run from $1$ to $2N$.  This group is isomorphic to $\text{Spin}(2N)$, the double cover of $SO(2N)$, whereas its action by conjugation on the operator algebra is $SO(2N)$.  The non-trivial element of $\text{Spin}(2N)$ which maps to the identity in $SO(2N)$ acts by $-1$ in the Fock space.  Now consider the subgroup $SU(N)\subset SO(2N)$ of those group elements that rotate the $N$ annihilation operators into each other, and let $|\psi\rangle$ be the all empty state.  The lift of this $SU(N)$ into $\text{Spin}(2N)$ consists of two disconnected components.  The component that contains the identity has the property that all operators $X$ in it satisfy $X|\psi\rangle = |\psi\rangle$, whereas all $X$ in the other component satisfy $X|\psi\rangle = -|\psi\rangle$.

Now suppose that the image of $V \in \text{Spin}(2N)$ in $SO(2N)$ sits inside this $SU(N)$.  Then there are two possibilities: either $V|\psi\rangle = |\psi\rangle$ or $V|\psi\rangle = -|\psi\rangle$.  How do we tell which is true?  To answer this, consider the image of the path $V(\tau)$ in $SO(2N)$, concatenated with some arbitrary path that connects $V(1)$ to $1$ within $SU(N) \subset SO(2N)$.  This is a loop in $SO(2N)$ that starts and ends at the identity, and since $SU(N)$ is simply connected, the element of $\pi_1(SO(2N))$ that it defines is independent of the choice of second path (the one within $SU(N)$).  Then, since $\text{Spin}(2N)$ is the double cover of $SO(2N)$, we will have $V|\psi\rangle = -|\psi\rangle$ if and only if this loop defines the non-trivial element of $\pi_1(SO(2N))$.

\subsubsection*{Action on Fock space for $U$ with winding number $1$}

Now let us consider our three dimensional system with two flavor bands $\nu$, and a free fermion unitary $U$ with non-trivial winding number $\nu_3(U)=1$.  Because the result of our computation will be quantized to $\pm 1$, we will obtain the same answer for any $U$ with $\nu_3(U)=1$, since they are all homotopic.  Rather than working with the $U$ we constructed in the main text, we will for convenience take the following specific form of $U$: $U_{\vec{k}} = \exp(i \pi f(|{\vec{k}}|){\vec{k}}\cdot {\vec{\sigma}})$.  Here $f(0)=1$ and $f(k) = 0$ for $k>k_0$, where $k_0$ is small compared to the inverse lattice spacing.  Also, $\vec{\sigma}$ is the vector of Pauli matrices in flavor space.  Hence $U$ is non-trivial only in a small neighborhood of $0$ around the Brillouin Zone (in particular, this is the only reflection symmetric point where $U\neq 1$).  This choice of $U$ clearly has a winding number of $1$, as the preimage of $-1\in SU(2)$ is just $\vec{k}=0$.  Note also that $U_{-\vec{k}} = U^\dagger_{\vec{k}}$.  As explained above, we first write $U$ as a unitary free fermion evolution $U(\tau)$, $0\leq \tau \leq 1$, with pairing, where $U(0)=1$ and $U(1)=U$.  Since $U$ preserves the all empty state, we can ask if its eigenvalue is $+1$ or $-1$ under this unitary free fermion evolution.  As discussed above, this will be a product of contributions over all pairs $({\vec{k}}, -{\vec{k}})$, times the product of contributions over the reflection-symmetric ${\vec{k}}$.  As explained above, each such contribution reduces to determining an element of a fundamental group defined by the path $U(\tau)$.

First, let us see that for non-reflection symmetric ${\vec{k}}$, the contribution from $({\vec{k}}, -{\vec{k}})$ to the eigenvalue is always $+1$.  This is simply because the pairing only takes place between ${\vec{k}}$ and $-{\vec{k}}$.  Thus, if we perform an anti-unitary particle-hole transformation which exchanges annihilation and creation operators at $-{\vec{k}}$ (but not at ${\vec{k}}$), then we map any such evolution to a particle number conserving unitary evolution; in this case it would take place in $SU(4)\subset SO(8)$.  Since $SU(4)$ is simply connected, by the discussion above the eigenvalue is $+1$.

We can be a little more explicit in our discussion of non-reflection symmetric ${\vec{k}}$.  Let us call the creation and annihilation operators over ${\vec{k}}$ $c_{1,2}, c_{1,2}^\dagger$, and those over $-{\vec{k}}$ $d_{1,2}, d_{1,2}^\dagger$.  Then suppose we have a free fermion Hamiltonian 

\begin{align*}
\sum_{i,j} H^{\veck}_{ij} c_i^\dagger c_j + \sum_{i,j} H^{-\veck}_{ij} c_i^\dagger c_j + \sum_{i,j} \Delta^\veck_{i,j} c_j d_i + \text{h. c.}
\end{align*}
where $H^{\veck}$ and $H^{-\veck}$ are traceless.  By doing an anti-unitary particle-hole transformation on $-\vec{k}$ which takes $d_i \leftrightarrow d_i^\dagger$, this maps to a particle conserving Hamiltonian whose matrix representation is 

\begin{align*}
\begin{pmatrix}
\left(H^{\veck}\right)^* & \Delta^\veck \\
\left(\Delta^\veck\right)^\dagger & - H^{-\veck}
\end{pmatrix}
\end{align*}
with the pairing term necessarily satisfying $\Delta^{-\veck} = - \left(\Delta^{\veck}\right)^T$.

Thus for reflection symmetric $\veck$, i.e. those satisfying $\veck = -\veck$,  $\Delta$ must be proportional to $\sigma^y$.  Letting $\vec{\tau}$ be the Pauli matrices on the $(\veck, -\veck)$ degree of freedom, we see that the possible free fermion Hamiltonians are then generated by $(\tau^z, \sigma^y\tau^y, \sigma^y\tau^x)$ and $(\sigma^x \tau^z, \sigma^y, \sigma^z \tau^z)$.  Note that these are two independent $su(2)$ Lie algebras; this is isomorphic to the Lie algebra of $SO(4)$.  This is what we expect: the allowed unitaries over the reflection symmetric $\veck$ must act as $SO(4)$ on the underlying Majoranas.

With this information in hand, let us now construct an explicit continuous path $U_\veck(\tau)$ connecting $1$ to $U_\veck$, for all $\veck$, and continuous in $\tau$ and $\veck$.  We will do this in the doubled framework, where we add particle-hole conjugates of the bands at $-\veck$ to the bands over each $\veck$.  Thus, we will connect the matrices 

\begin{align*}
{\bf{U}}_\veck = \begin{pmatrix}
\exp(-i \pi f(|{\vec{k}}|){\vec{k}}\cdot {\vec{\sigma}}) & 0 \\
0 & \exp(i \pi f(|{\vec{k}}|){\vec{k}}\cdot {\vec{\sigma}})
\end{pmatrix}
\end{align*}
to the identity, continuously in $\veck$ and respecting the $\veck \leftrightarrow -\veck$ particle-hole redundancy.  To do this, first consider the path in $\theta$, for $0\leq \theta \leq \pi/2$:

\begin{align*}
{\bf{U}}_\veck(\theta) = \cos(\theta){\bf{U}}_\veck + \sin(\theta)\sigma^y \tau^y
\end{align*}
Then simply concatenate this with the reversal of the path

\begin{align*}
\cos(\theta)\cdot 1 + \sin(\theta)\sigma^y \tau^y
\end{align*}
$0\leq \theta \leq \pi/2$.  Note that for $|\veck| > k_0$, these two paths are just reverses of each other, so there they can be homotoped into the identity path.  Let us perform this homotopy, so that our thus modified path is non-trivial only for $|\veck| \leq k_0$.

Now, at $\veck=0$, this path is something that connects $U_{\veck=0}=-1$ to $1$ entirely in the $SU(2)$ Lie group generated by $(\tau^z, \sigma^y\tau^y, \sigma^y\tau^x)$.  Noting that the quotient of $SO(4)$ by the other $SU(2)$ Lie group (the one generated by $(\sigma^x \tau^z, \sigma^y, \sigma^z \tau^z)$) is equal to this $SU(2)$ divided by its center (which the two $SU(2)$'s in $SO(4)$ share in common), we see that this gives a non-trivial loop in $SO(3) = SO(4)/SU(2)$.  The $SU(2)$ in the denominator of this quotient is the one that preserves the all empty state, so considering this path in $SO(4)$ and then connecting back to the identity through the $SU(2)$ in the denominator gives a closed non-trivial loop in $SO(4)$ (otherwise the loop in $SO(3)$ would have been trivial).  So by the discussion above, we get a factor of $-1$ from the action of the unitary evolution at $\veck = 0$.  This is the only non-trivial contribution, so we see that $U$ acts as $-1$ on Fock space when all three boundary conditions are anti-periodic, as was to be shown.

One might ask to what extent our result depends on the particular circuit chosen to represent $U$.  Two different such circuits differ by a loop in the space of local free fermion unitary evolutions, which is known to be trivial in $3$ spatial dimensions \cite{harper}.  Now there are loops in the space of $2d$ and $1d$ free fermion unitaries, namely those found by Rudner et al.\cite{RudnerLevin}, and the Thouless charge pump. These can give rise to $CZ$ gates along various $T^2$'s in the $T^3$, or $Z$ gates along various $S^1$'s in $T^3$, respectively.  But they cannot eliminate the top dimensional invariant, which is the $CCZ$ gate.

%\subsubsection*{Proof that the $SU(2) \subset SO(4)$ mentioned above preserves the all empty state}

%Let's try to be a bit more explicit about the reflection invariant points.  Let $a,a^\dag, b, b^\dag$ represent the two bands.  In a Majorana notation we have

%\begin{align*}
%a &= (\gamma_1+i\gamma_2)/2 \\
%a^\dag &= (\gamma_1 - i\gamma_2)/2 \\
%b &= (\gamma_3+i\gamma_4)/2 \\
%b^\dag &= (\gamma_3 - i\gamma_4)/2
%\end{align*}

%Then the following is true:

%\begin{align*}
%a b^\dag + b a^\dag &= (-i \gamma_1 \gamma_4 + i \gamma_2 \gamma_3)/2 \\
%i(a b^\dag - b a^\dag) &= (i \gamma_1 \gamma_3 + i \gamma_2 \gamma_4)/2 \\
%a a^\dag - b b^\dag &= (-i \gamma_1 \gamma_2 + i \gamma_3 \gamma_4)/2
%\end{align*}
%These are generators of $su(2)$ inside $so(4)$.  This $su(2)$ preserves the all empty state.  As far as Lie groups, these generate $SU(2)$ inside $SO(4)$, and $SO(4)/SU(2)=SO(3)$.  This comes from the fact that this $SU(2)$ is one factor of $SO(4)=SU(2) \times SU(2) / {\mathbb{Z}_2}$.

\bibliography{pump-ref.bib}

\end{document}